\documentclass[aps,twocolumn,showpacs,amsmath,amssymb,superscriptaddress,prc]{revtex4-2}
\usepackage{graphicx,here}
\usepackage{multirow}
\usepackage{tikz}
\usepackage{epstopdf}
\usepackage[percent]{overpic}
\usepackage{scrextend}
\usepackage{xcolor,xspace}
\usepackage[ulem=normalem]{changes}
\usepackage{hyperref}
\hypersetup{colorlinks=True,urlcolor=blue,linkcolor=blue,citecolor=blue,filecolor=black}

\colorlet{Changes@Color}{red}
\usepackage{dcolumn,color,footnote,bm,braket}
\usepackage{url,longtable,tabularx}
\usepackage{threeparttable}

\usepackage{fancybox}
\usetikzlibrary{matrix}

\begin{document}

\title{Microscopic description of hexadecapole collectivity in even-even rare-earth nuclei near $N=90$}

\author{L. Lotina}
\email{llotina.phy@pmf.hr}
\affiliation{Department of Physics, Faculty of Science, 
University of Zagreb, HR-10000 Zagreb, Croatia}

\author{K. Nomura}
\email{nomura@sci.hokudai.ac.jp}
\affiliation{Department of Physics, 
Hokkaido University, Sapporo 060-0810, Japan}
\affiliation{Nuclear Reaction Data Center, 
Hokkaido University, Sapporo 060-0810, Japan}

\date{\today}

\begin{abstract}
We present an extensive study of hexadecapole correlations in the rare-earth region near $N=90$ and the effects these correlations have on various nuclear properties, such as the low-energy spectra, as well as quadrupole, hexadecapole, 
and monopole transition strengths. In order to examine hexadecapole correlations, we employ a mapped $sdg$ interacting boson model, with parameters derived from a self-consistent mean-field calculations with a relativistic energy density functional. We apply this model to even-even isotopes 
of Nd, Sm, Gd, Dy, and Er ($Z=60 - 68$) with neutron numbers $N=84-96$. 
The obtained results show a good agreement with the experiment. By comparing the results with the ones obtained from a simpler mapped $sd$ interacting boson model, we show that the inclusion of the hexadecapole 
degree of freedom via $g$ boson 
is necessary to improve the results of the $J^{\pi} \geq 6^{+}$ yrast energies in the nuclei with 
$N=84$ and 86, being near the neutron shell closure. 
The $sdg$ interacting boson model increases 
the quadrupole transition strengths between yrast states 
in the $N=90$ and 92 well deformed nuclei, 
which is in good agreement with the experiment for most of those isotopes. 
The presence of $g$ bosons does have an important effect 
on hexadecapole transition strengths, although experimental 
data for such transitions are limited. 
The obtained monopole transition strengths do not differ significantly from the ones obtained from the simpler $sd$ model.
\end{abstract}

\maketitle

\section{Introduction}

Nuclear deformations play an important role in describing various nuclear properties \cite{BM,RS}, e.g., 
excitation energies and decays. 
The dominant deformations in nuclei, the quadrupole ones, 
have been extensively studied. More and more attention is recently 
being paid to higher-order deformations, such as the octupole 
and hexadecapole ones. 
The effects of hexadecapole correlations are often overshadowed by large quadrupole effects. Nevertheless, they have been found to exist in a wide spectrum of nuclei, ranging from light nuclei \cite{gupta2020} to heavy nuclei \cite{ryssens2023}. The main effect of hexadecapole correlations on the low-lying energy spectrum of the nucleus is the appearance of the low-lying $K=4^+$ band with an enhanced $B(E4; 4^+ \rightarrow 0^+)$ transition strength. Another effect can be observed in even-even rare-earth nuclei near the 
$N=82$ shell closure, where the ratio of the ground-state band energies $R_{4/2} = E_{x}(4^+)/E_{x}(2^+)$ becomes less than 2. Besides that, hexadecapole deformations were shown to play a significant role in heavy ion collisions \cite{ryssens2023} and fission \cite{chi2023}, 
and are predicted to have an influence on the neutrinoless double beta decay matrix elements in open shell nuclei \cite{engel2017}. All of this provides us with a good reason to study hexadecapole correlations in nuclei and their effects on the low-lying excitation spectra and transitions.

A useful framework for studying the effects of nuclear deformations is the interacting boson model (IBM) \cite{IBM}. In the simplest version of the IBM, the nucleus can be viewed as a system composed of a 
doubly magic core nucleus, and valence nucleons grouped into $s$ ($L^{\pi}=0^+$) and $d$ $(L^{\pi}=2^+)$ bosons. The main assumption of the model is that the main contribution to the low-lying excitation energy spectra comes from the pairing correlations between aforementioned bosons. In the version of the model called IBM-1, it is assumed that the neutron and proton bosons are identical \cite{IBM}. This model has been successfully used to study the effects of deformations in nuclei \cite{casten1988}. Since the IBM is a phenomenological model, in recent times, a method was developed that derives the parameters of the IBM Hamiltonian from a self-consistent mean-field (SCMF) model with energy density functionals (EDFs) \cite{nomura2008}. This method has been successfully applied in studying quadrupole \cite{nomura2008, nomura2010, nomura2011pt, nomura2012tri} and octupole correlations \cite{nomura2013oct, nomura2014} in nuclei. The inclusion of the hexadecapole degree of freedom in the IBM is done by 
the inclusion of the $g$ boson 
with $L^{\pi}=4^+$, whose importance in the IBM 
has been extensively studied 
\cite{otsuka1981, otsuka1982, otsuka1985, otsuka1988, casten1988, devi-kota1990, kuyucak1994, vanisacker2010}. 
While the $sdg$-IBM has been extensively 
studied as a phenomenological model, 
it is useful to study the model through 
a more microscopic picture, 
e.g., the aforementioned mapping method, 
since that could lead us to a better 
understanding of the microscopic origin 
of the hexadecapole collectivity in nuclei. 

In the preceding article \cite{lotina2024}, 
we explored the hexadecapole collectivity in $^{148-160} \textnormal{Gd}$ isotopes by using the $sdg$-IBM-1 with the Hamiltonian parameters being derived by the mapping method of Ref.~\cite{nomura2008}, and 
showed the validity and usefulness of such approach. 
The aim of the present article 
is to extend the study to a wider range of even-even rare-earth isotopes, $^{144-156}\textnormal{Nd}$, $^{146-158}\textnormal{Sm}$, $^{148-160}\textnormal{Gd}$, $^{150-162}\textnormal{Dy}$, 
and $^{152-164}\textnormal{Er}$. We choose the rare-earth isotopes for our study due to the fact that hexadecapole correlations were observed in that region \cite{BM, hendrie1968, erb1972, wollersheim1977, ronningen1977, ronningen21977}, as well as due to the fact that triaxiality does not play a significant role in these isotopes, as is evident from SCMF calculations with the Skyrme force \cite{nomura2010} and Gogny force \cite{hilaire2007}. By comparing our model with a simpler mapped $sd$-IBM-1, we explore the effects of hexadecapole correlations on the low-lying excitation energy spectra of these isotopes, and also on the monopole, quadrupole, and hexadecapole transition strengths. 

The paper is organized as follows. 
In Sec.~\ref{sec:model} we describe our model. 
Section~\ref{sec:pes} gives the quadrupole-hexadecapole 
potential energy surfaces for the studied nuclei. 
Results of the spectroscopic properties, 
including the excitation spectra of low-lying 
states and the electric quadrupole, hexadecapole, 
and monopole transition properties, 
are discussed in Sec.~\ref{sec:results}. 
A summary of the main results and some 
perspectives for future work are 
given in Sec.~\ref{sec:summary}.

%-----------------------------------------------------------
%      Model description
%-----------------------------------------------------------

\section{Model description\label{sec:model}}

We begin our analysis with the SCMF calculations. 
The model employed for SCMF calculations is the 
multidimensionally constrained relativistic mean-field (MDC-RMF) model \cite{zhou2016, lu2014, zhao2017}, which allows one to set constraints on various deformation parameters. For our analysis, we carried out the SCMF calculations for axially symmetric shapes in the ($\beta_2, \beta_4$) plane, by setting the constraints on the mass quadrupole $Q_{20}$ and hexadecapole $Q_{40}$ moments. The dimensionless quadrupole $\beta_2$ and hexadecapole $\beta_4$ deformation parameters are related to the mass moments through the relation
\begin{equation}
    \beta_{\lambda}= \frac{4\pi }{3AR^{\lambda }} \braket{\hat{Q}_{\lambda 0}}
    \label{eq1}
\end{equation}
with $R=1.2 A^{1/3}$ fm. 
The quadrupole-hexadecapole 
constrained potential energy surfaces (PESs) 
are calculated within the relativistic Hartree-Bogoliubov (RHB) 
framework \cite{vretenar2005, niksic2011}, 
with the chosen energy density functional 
being the density-dependent point-coupling (DD-PC1) 
interaction \cite{DDPC1, niksic2011}, combined with 
the separable pairing interaction of finite range 
developed in Ref.~\cite{tian2009}. 
A detailed description of the MDC-RMF model 
can be found in Refs.~\cite{lu2014,zhao2017}.

Due to the fact that SCMF calculations necessarily break several symmetries, these calculations alone cannot be used to study excited states and transitions in the nucleus. To study those properties of the nucleus, we use the $sdg$-IBM-1 model. A simple version of the $sdg$-IBM-1 Hamiltonian is given by the following relation, similar to the one from \cite{kuyucak1994}:
\begin{equation}
    \hat{H}_{sdg}= \epsilon_d \hat{n}_d + \epsilon_g \hat{n}_g + \kappa_2 \hat{Q}^{(2)} \cdot \hat{Q}^{(2)} + \kappa_4 \hat{Q}^{(4)} \cdot \hat{Q}^{(4)}.
    \label{eq2}
\end{equation}
The first two terms represent the $d$ and $g$ boson number operators, $\hat{n}_d = d^{\dagger} \cdot \tilde{d}$ and $\hat{n}_g = g^{\dagger} \cdot \tilde{g}$. The second term represents the quadrupole-quadrupole interaction with the quadrupole operator, defined as
\begin{equation}
\begin{aligned}
     \hat{Q}^{(2)} = &(s^{\dagger} \tilde{d} + d^{\dagger} s) + \chi^{(2)}_{dd}  (d^{\dagger} \times \tilde{d})^{(2)} \\ 
     & + \chi^{(2)}_{dg} (d^{\dagger} \times \tilde{g} + g^{\dagger} \times \tilde{d})^{(2)} + \chi^{(2)}_{gg} (g^{\dagger} \times \tilde{g})^{(2)},
\end{aligned}
\label{eq3}
\end{equation}
while the last term represents the hexadecapole-hexadecapole interaction, with the hexadecapole operator defined as:
\begin{equation}
 \begin{aligned}
     \hat{Q}^{(4)} = &(s^{\dagger} \tilde{g} + g^{\dagger} s) + \chi^{(4)}_{dd}  (d^{\dagger} \times \tilde{d})^{(4)} \\ 
     & + \chi^{(4)}_{dg} (d^{\dagger} \times \tilde{g} + g^{\dagger} \times \tilde{d})^{(4)} + \chi^{(4)}_{gg} (g^{\dagger} \times \tilde{g})^{(4)}.
\end{aligned}
\label{eq4}
\end{equation}
Since this Hamiltonian is too complex, due to the number of parameters it contains, a simplification can be made by assuming three symmetry limits, U(5) $\otimes$ U(9), SU(3), and SO(15), which leads to a Hamiltonian \cite{vanisacker2010}
\begin{equation}
    \hat H_{sdg} = \epsilon_d \hat n_d + \epsilon_g \hat n_g + \kappa \hat{Q}^{(2)} \cdot \hat{Q}^{(2)} + \kappa (1-\chi^2) \hat{Q}^{(4)} \cdot \hat{Q}^{(4)} 
    \label{eq5}
\end{equation}
with 
\begin{equation}
\begin{aligned}
  \hat{Q}^{(2)} = &(s^{\dagger} \tilde{d} + d^{\dagger} s) + \chi  \Big [ \frac{11 \sqrt{10}}{28}(d^{\dagger} \times \tilde{d})^{(2)} \\
      & - \frac{9}{7} (d^{\dagger} \times \tilde{g} + g^{\dagger} \times \tilde{g})^{(2)} + \frac{3 \sqrt{55}}{14} (g^{\dagger} \times \tilde{g})^{(2)} \Big ] 
\end{aligned}
\label{eq6}
\end{equation}
and
\begin{equation}
    \hat{Q}^{(4)} = s^{\dagger} \tilde{g} + g^{\dagger} s
    \label{eq7}
\end{equation}
being the quadrupole and hexadecapole operators, respectively.

The parameters $\epsilon_d$, $\epsilon_g$, $\kappa$, 
and $\chi$ are determined by the mapping procedure \cite{nomura2008}. The first step is connecting the IBM to the geometric model by calculating the expectation value of the Hamiltonian in a coherent state $\ket{\phi} \propto (1+ \tilde{\beta}_2 d^{\dagger}_0 + \tilde{\beta}_4 g^{\dagger}_0)^{N_B} \ket{0}$, with $N_B$ representing the number of bosons, i.e., the number 
of pairs of valence nucleons, and $\ket{0}$ representing 
the boson vacuum \cite{vanisacker2010}. 
For Nd, Sm, Gd and Dy isotopes, the boson vacuum 
corresponds to the double shell closures 
($N,Z$) = (82, 50), i.e., the doubly magic nucleus $^{132}$Sn, 
while for the Er isotopes, since the valence neutrons 
are considered hole-like, the corresponding 
boson vacuum is taken to be 
($N,Z$)=(82, 82). 
The expectation value, 
$\braket{\phi | \hat H | \phi}/\braket{\phi|\phi}$, 
gives us the PES of the IBM, and is 
denoted $E_{\textnormal{IBM}}(\tilde{\beta}_2, \tilde{\beta}_4)$, 
with $\tilde{\beta}_2$ and $\tilde{\beta}_4$ being 
boson analogs of the quadrupole $\beta_2$ 
and $\beta_4$ deformations, respectively. 
The parameters of the Hamiltonian 
are fitted so that the energy surface of the IBM approximates 
the PES obtained from the SCMF calculations, 
$E_{\textnormal{SCMF}}(\beta_2, \beta_4)$, 
in the vicinity of the minimum:
\begin{equation}
    E_{\textnormal{SCMF}}(\beta_2, \beta_4) \approx E_{\textnormal{IBM}}(\tilde{\beta}_2, \tilde{\beta}_4).
    \label{eq8}
\end{equation}
Following the method of Refs. \cite{nomura2008, nomura2014}, 
the relation between bosonic and fermionic deformation 
parameters is assumed to be linear, 
$\tilde{\beta}_2 = C_2 \beta_2$, $\tilde{\beta}_4= C_4 \beta_4$. This leaves us with six parameters in total to be determined. 
In the case of lighter rare-earth isotopes, Nd and Sm, the Hamiltonian from Eq. (\ref{eq6}) is shown to be inadequate to reproduce the SCMF PES, due to the obtained ratios $\beta_{4}^{\textnormal{min}}/\beta_{2}^{\textnormal{min}}$ being larger than in heavier rare-earth isotopes. To solve this problem, an independent parameter $\sigma$ was introduced in the quadrupole operator of Eq. (\ref{eq6}) as:
\begin{equation}
\begin{aligned}
  \hat{Q}^{(2)} = &(s^{\dagger} \tilde{d} + d^{\dagger} s) + \chi  \Big [ \frac{11 \sqrt{10}}{28}(d^{\dagger} \times \tilde{d})^{(2)} \\
      & - \frac{9}{7} \sigma (d^{\dagger} \times \tilde{g} + g^{\dagger} \times \tilde{g})^{(2)} + \frac{3 \sqrt{55}}{14} (g^{\dagger} \times \tilde{g})^{(2)} \Big ] ,
\end{aligned}
\label{eq9}
\end{equation} 
with constraint $-1 \leq \chi \sigma \leq +1$. 
If $\chi = \sigma = +1$, the quadrupole operator corresponds to the generator of the SU(3) algebra \cite{kota1987}. 
It should be noted that, while the hexadecapole terms and the $(g^{\dagger} \times \tilde{g})^{(2)}$ are included in the Hamiltonian, their contribution to the IBM PES, as well as to the excitation energies, is minimal, and they could, in principle, be omitted from the Hamiltonian.

In order to study the effects of hexadecapole correlations in nuclei, the $sdg$ IBM has to be compared with a simpler $sd$ IBM, with a Hamiltonian given by the relation \cite{IBM}
\begin{equation}
    \hat{H}_{sd} = \epsilon_d \hat{n}_d + \kappa \hat{Q}^{(2)} \cdot \hat{Q}^{(2)},
    \label{eq10}
\end{equation}
with 
\begin{equation}
    \hat{Q}^{(2)}= s^{\dagger} \tilde{d} + d^{\dagger} s + \chi(d^{\dagger} \times \tilde{d})^{(2)}
    \label{eq11}
\end{equation}
being the quadrupole operator. The mapping is performed so that the energy of the $sd$ IBM approximates the SCMF PES along the $\beta_4=0$ line in the vicinity of the minimum \cite{lotina2024}:
\begin{equation}
    E_{\textnormal{SCMF}}(\beta_2, \beta_4=0) \approx E_{\textnormal{IBM}}(\tilde{\beta}_2),
    \label{eq12}
\end{equation}
while the relation between the bosonic and fermionic quadrupole deformation parameters is again assumed to be linear, $\tilde{\beta}_2 = C_2^{sd}\beta_2$.

The transition strengths are defined as
\begin{equation}
    B(E \lambda ; J \rightarrow J') = \frac{1}{2J + 1}  | \bra{J'}  | \hat{T}^{(E \lambda)} | \ket{J} |^2,
    \label{eq13}
\end{equation}
with $\ket{J}$ and $\bra{J'}$ being the wave functions of the initial and final states, respectively. The operators considered are the quadrupole operator
\begin{equation}
    \hat{T}^{(E2)} = e_2^{sdg, sd} \hat{Q}^{(2)},
    \label{eq14}
\end{equation}
with $\hat{Q}^{(2)}$ corresponding to the quadrupole operator of the $sdg$- or $sd$-IBM [Eqs. (\ref{eq2}), (\ref{eq6}), and (\ref{eq10})]; 
the hexadecapole operator, defined as
\begin{equation}
    \hat{T}^{(E4)} = e_4^{sdg} \left [ s^{\dagger} \tilde{g} + g^{\dagger} s + (d^{\dagger} \times \tilde{d})^{(4)} \right ]
    \label{eq15}
\end{equation}
for the $sdg$-IBM, and
\begin{equation}
  \hat{T}^{(E4)} = e_4^{sd} (d^{\dagger} \times \tilde{d})^{(4)}
  \label{eq16}
\end{equation}
for the $sd$-IBM; 
and the monopole operator, defined as \cite{vanisacker2012}
\begin{equation}
    \hat{T}^{(E0)} = (e_n N + e_p Z) \left (\eta \frac{\hat{n}_d}{N_B} + \gamma \frac{\hat{n}_g}{N_B} \right)
    \label{eq17}
\end{equation}
for the $sdg$-IBM and
\begin{equation}
   \hat{T}^{(E0)} = (e_n N + e_p Z) \eta \frac{\hat{n}_d}{N_B}
   \label{eq18}
\end{equation}
for the $sd$-IBM. The $e_2^{sdg, sd}$ coefficients are fitted for each isotope so that the experimentally measured transition strength $B(E2; 2_1^+ \rightarrow 0_1^+)$ from the first $2^+$ state to the ground state should be reproduced. Similarly, the $e_4^{sdg, sd}$ coefficients are fitted to the $B(E4; 4_1^+ \rightarrow 0_1^+)$ transition strength. In the case of monopole transitions, following Ref. \cite{vanisacker2012}, monopole strengths are defined as
\begin{equation}
    \rho(E0) = \frac{\bra{J'} \hat{T}^{(E0)} \ket{J}}{eR^2},
    \label{eq19}
\end{equation}
with $R=1.2 A^{1/3} \textnormal{fm}$ being the nuclear radius. 
The parameters $e_{p,n}$ are chosen to be $e_n = 0.50 e$, $e_p = e$, 
following Ref. \cite{vanisacker2012}. 
However, a different choice from the one 
in Ref.~\cite{vanisacker2012} is made for 
these parameters, 
$\eta = \gamma = 0.75 \: \textnormal{fm}^2$. 
This is due to the fact that the Hamiltonians 
used in this paper are different from ones used in the aforementioned paper. Most of the experimental data are taken from the National Nuclear Data Center (NNDC) database \cite{data}.

%-----------------------------------------------------------
%
%       PES figures, SCMF and IBM calculations
%
%-----------------------------------------------------------
\begin{figure*}    
\begin{center}
\includegraphics[width = 0.85\textwidth]{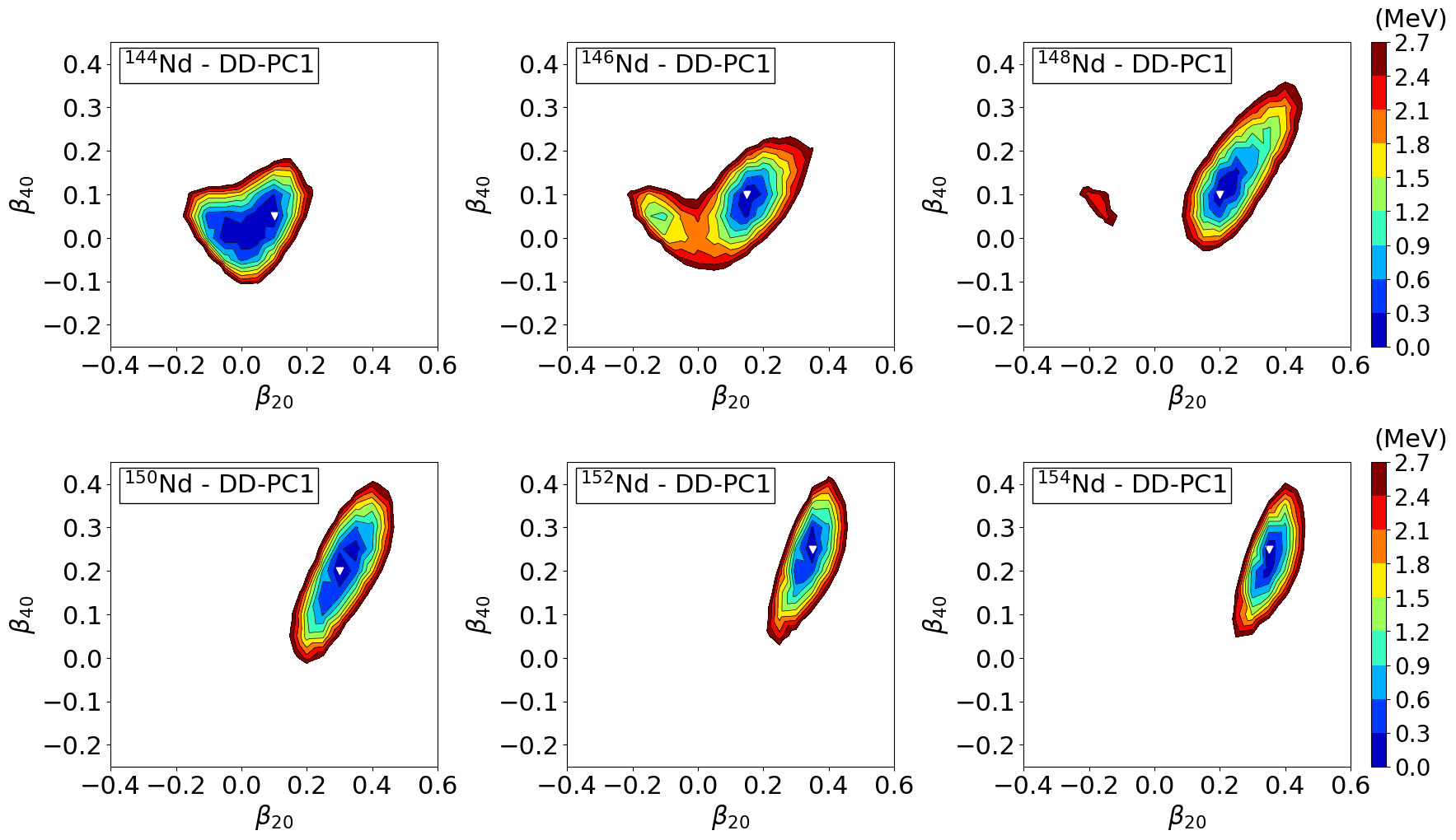}
\caption{Axially symmetric quadrupole ($\beta_{20}$) and 
hexadecapole ($\beta_{40}$) 
constrained energy surfaces for the $^{144-154}$Nd 
isotopes calculated within the relativistic Hartree-Bogoliubov 
method using the DD-PC1 energy density functional 
and the pairing force of finite range. 
Energy difference between neighboring contours is 
0.3 MeV, and the absolute minimum is indicated by 
an open triangle.} 
\label{NDSCMF}
\end{center}
\end{figure*}

\begin{figure*}    
\begin{center}
\includegraphics[width = 0.85\textwidth]{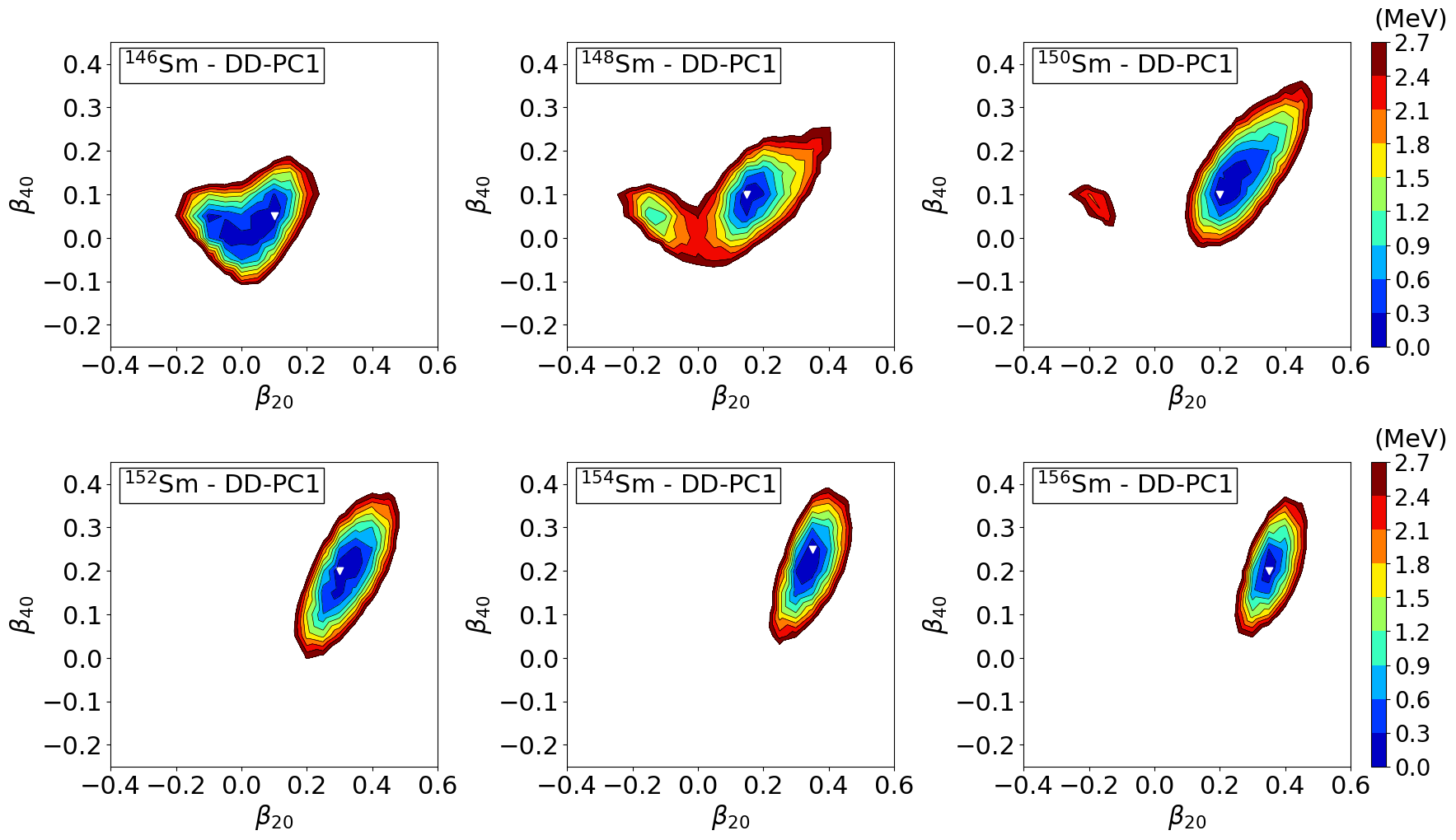}
\caption{Same as the caption for Fig. \ref{NDSCMF}, but for $^{146-156}$Sm} 
\label{SMSCMF}
\end{center}
\end{figure*}

\begin{figure*}    
\begin{center}
\includegraphics[width = 0.85\textwidth]{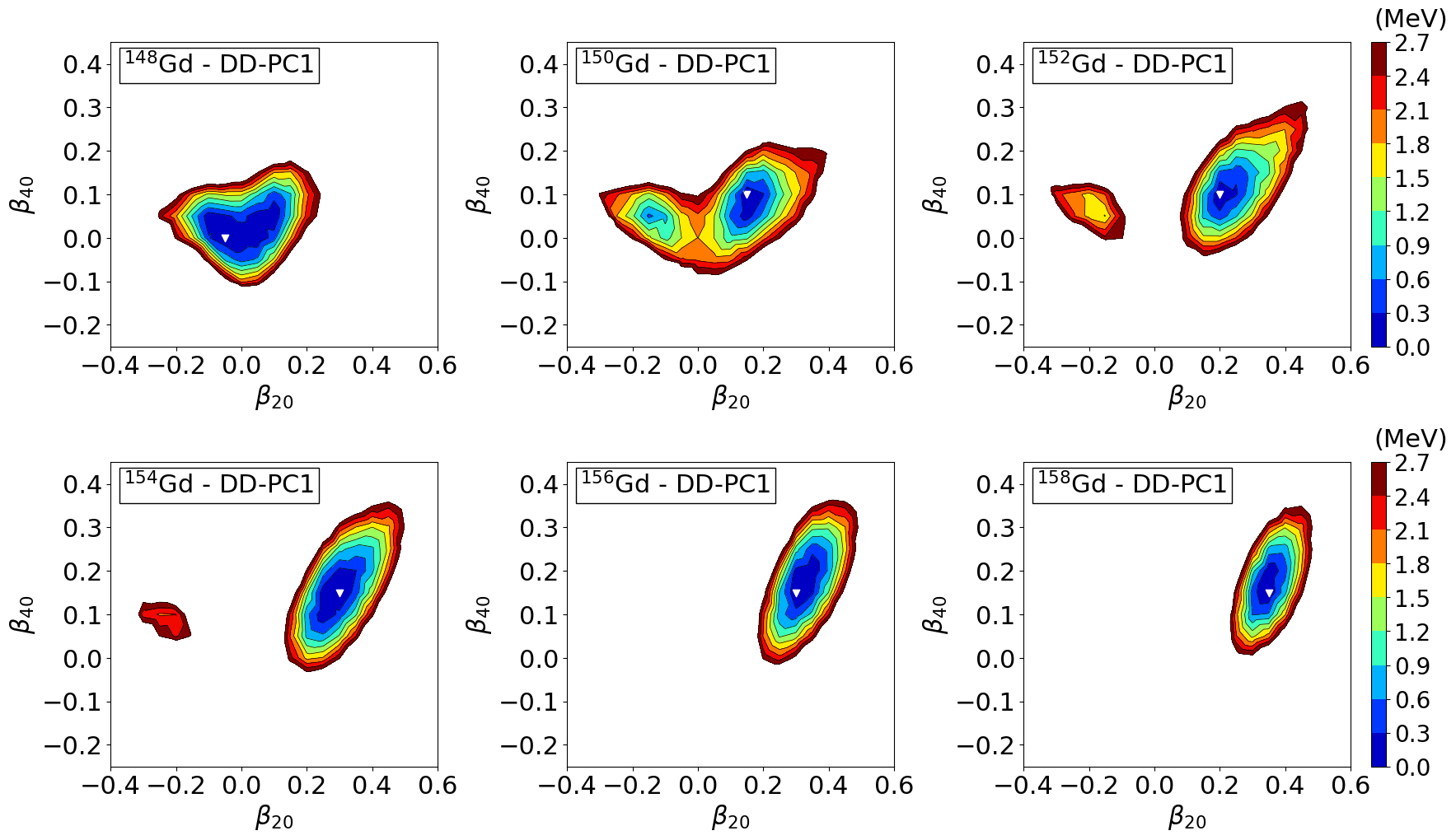}
\caption{Same as the caption for Fig. \ref{NDSCMF}, but for $^{148-158}$Gd. } 
\label{GDSCMF}
\end{center}
\end{figure*}

\begin{figure*}    
\begin{center}
\includegraphics[width = 0.85\textwidth]{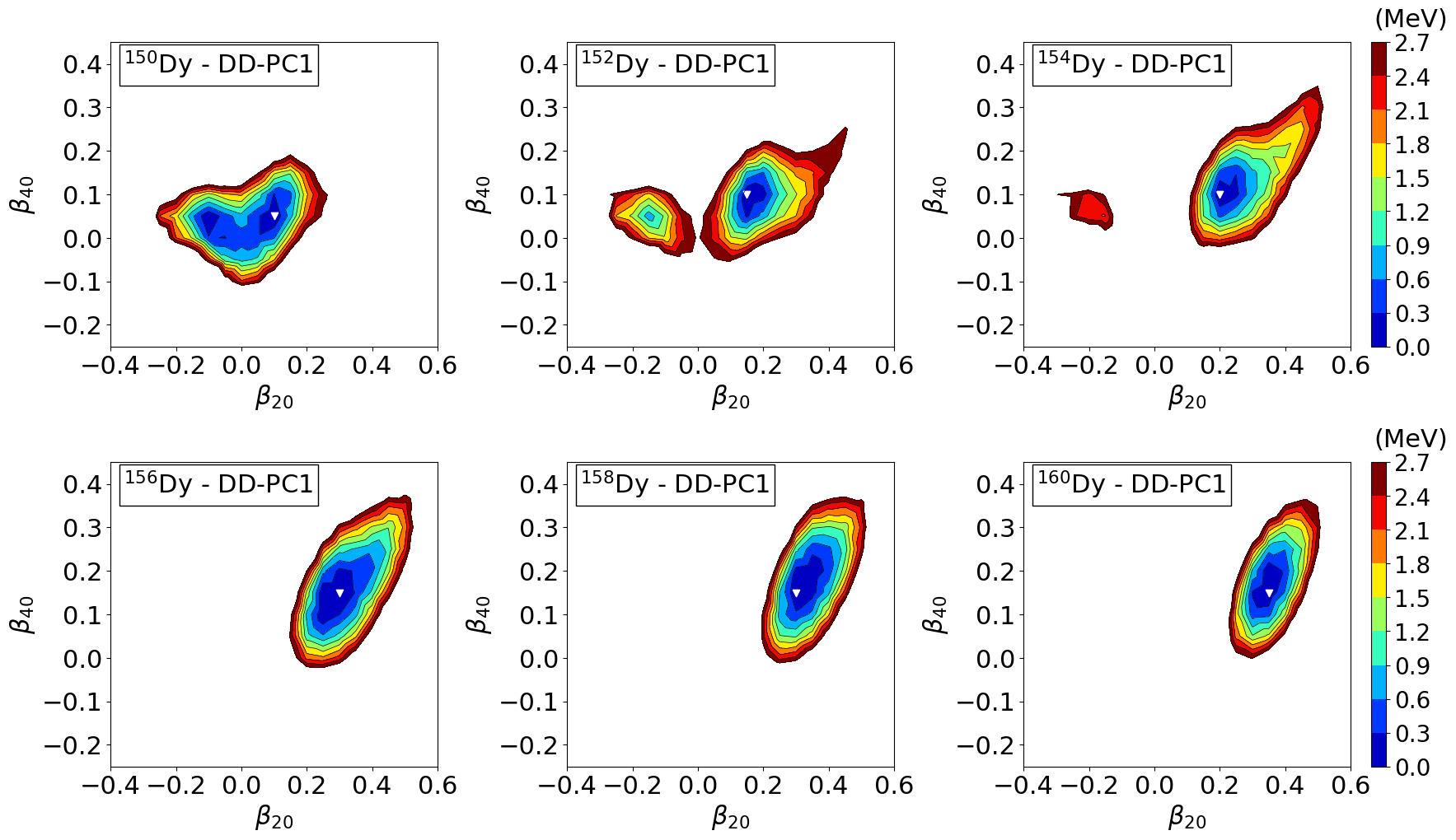}
\caption{Same as the caption for Fig. \ref{NDSCMF}, but for $^{150-160}$Dy} 
\label{DYSCMF}
\end{center}
\end{figure*}

\begin{figure*}    
\begin{center}
\includegraphics[width = 0.85\textwidth]{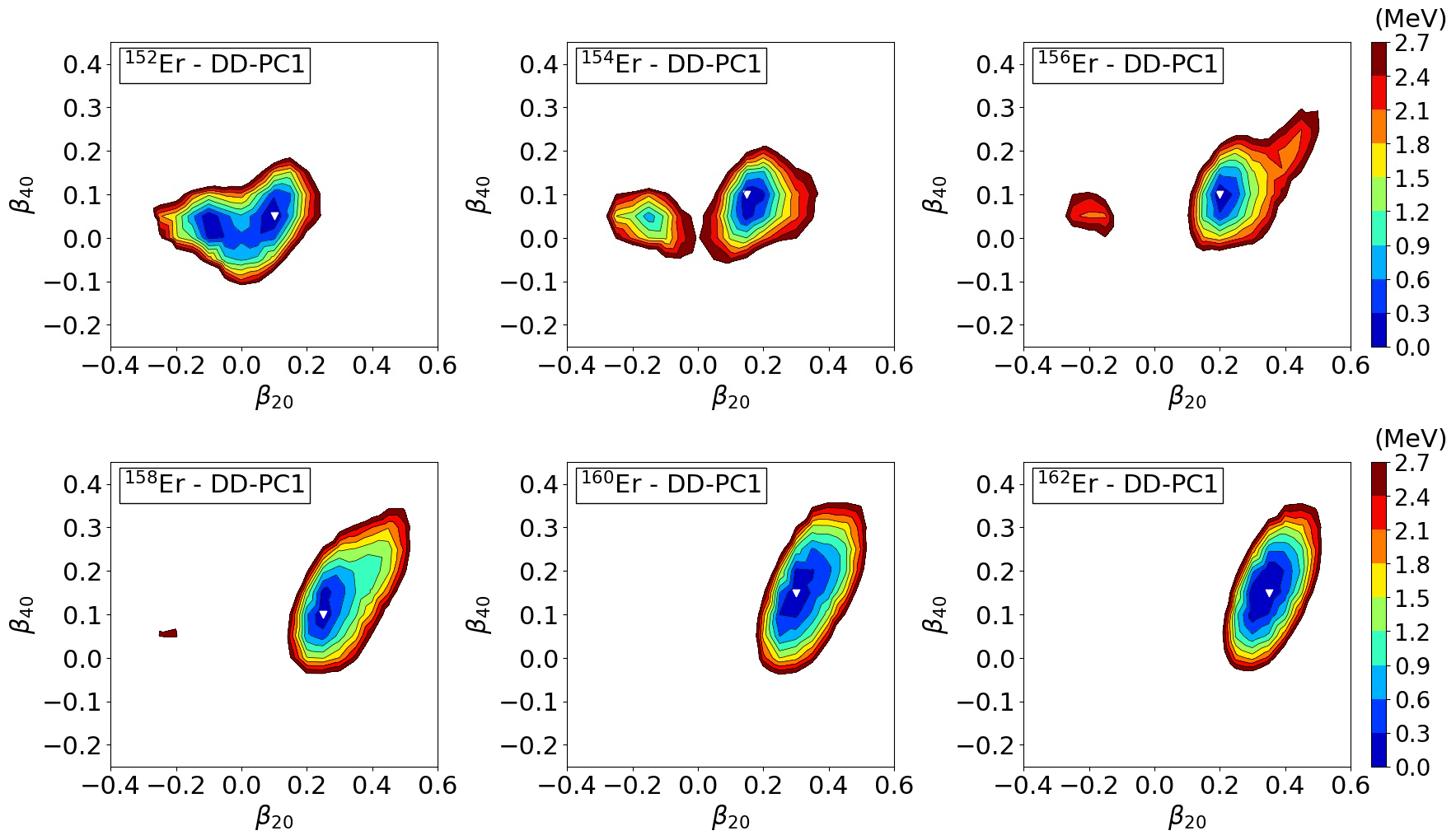}
\caption{Same as the caption for Fig. \ref{NDSCMF}, but for $^{152-162}$Er} 
\label{ERSCMF}
\end{center}
\end{figure*}

\begin{figure*}    
\begin{center}
\includegraphics[width = 0.85\textwidth]{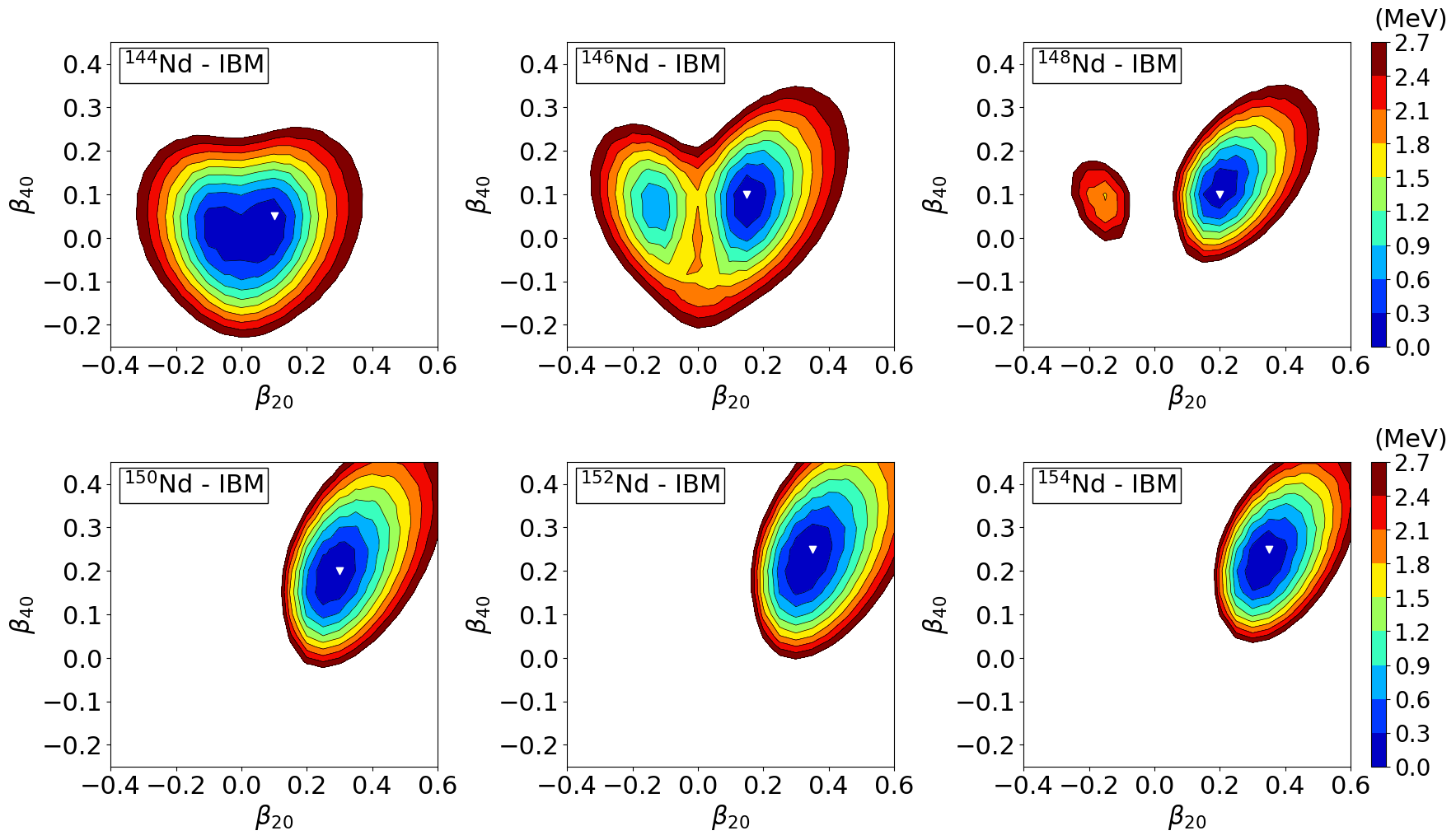}
\caption{Same as the caption for Fig. \ref{NDSCMF}, but for the mapped $sdg$-IBM energy surfaces of $^{144-154}$Nd.} 
\label{NDIBM}
\end{center}
\end{figure*}

\begin{figure*}    
\begin{center}
\includegraphics[width = 0.85\textwidth]{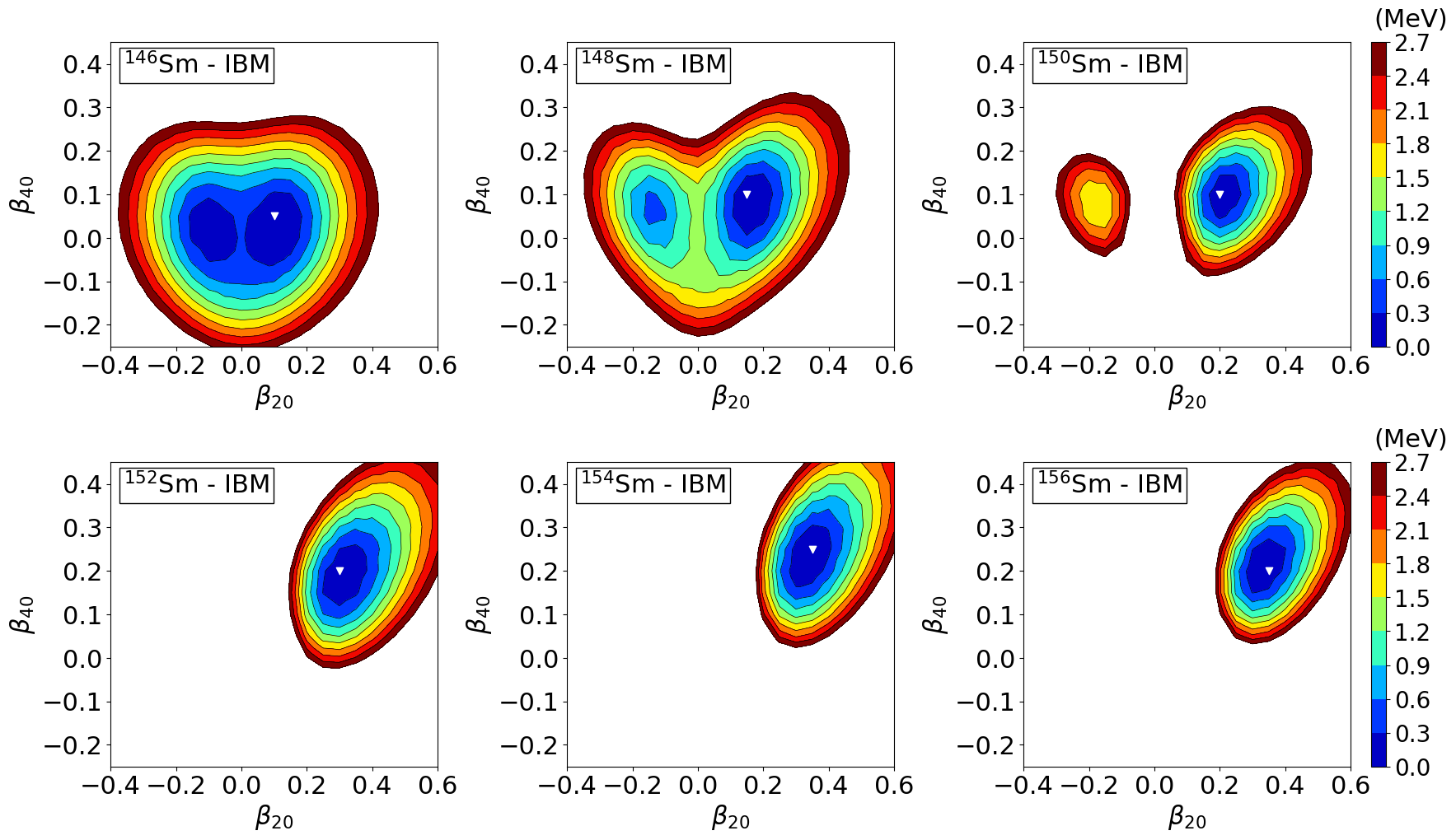}
\caption{Same as the caption for Fig. \ref{NDSCMF}, but for the mapped $sdg$-IBM energy surfaces of $^{146-156}$Sm.} 
\label{SMIBM}
\end{center}
\end{figure*}

\begin{figure*}    
\begin{center}
\includegraphics[width = 0.85\textwidth]{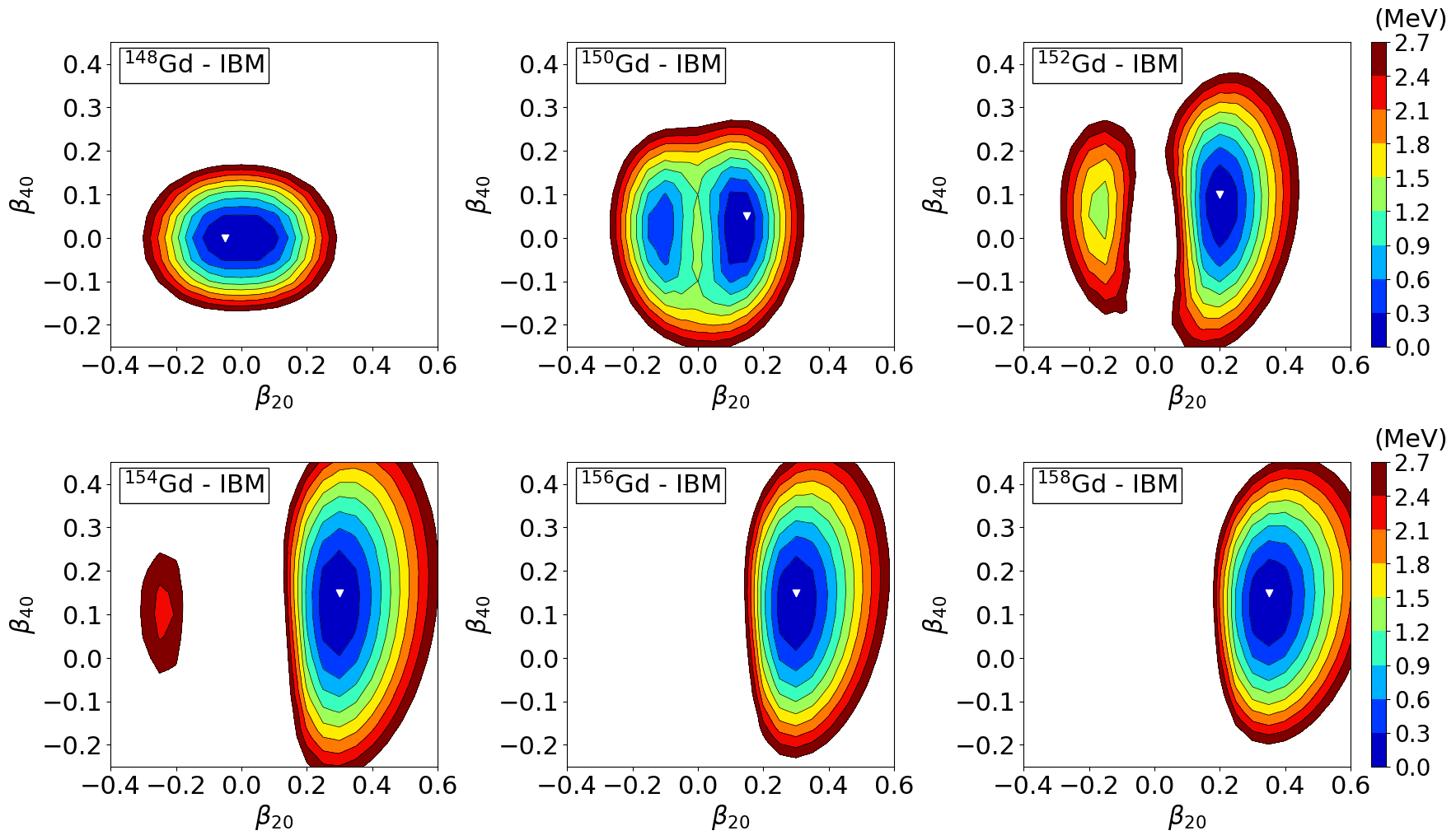}
\caption{Same as the caption for Fig. \ref{NDSCMF}, but for the mapped $sdg$-IBM energy surfaces of $^{148-158}$Gd.} 
\label{GDIBM}
\end{center}
\end{figure*}

\begin{figure*}    
\begin{center}
\includegraphics[width = 0.85\textwidth]{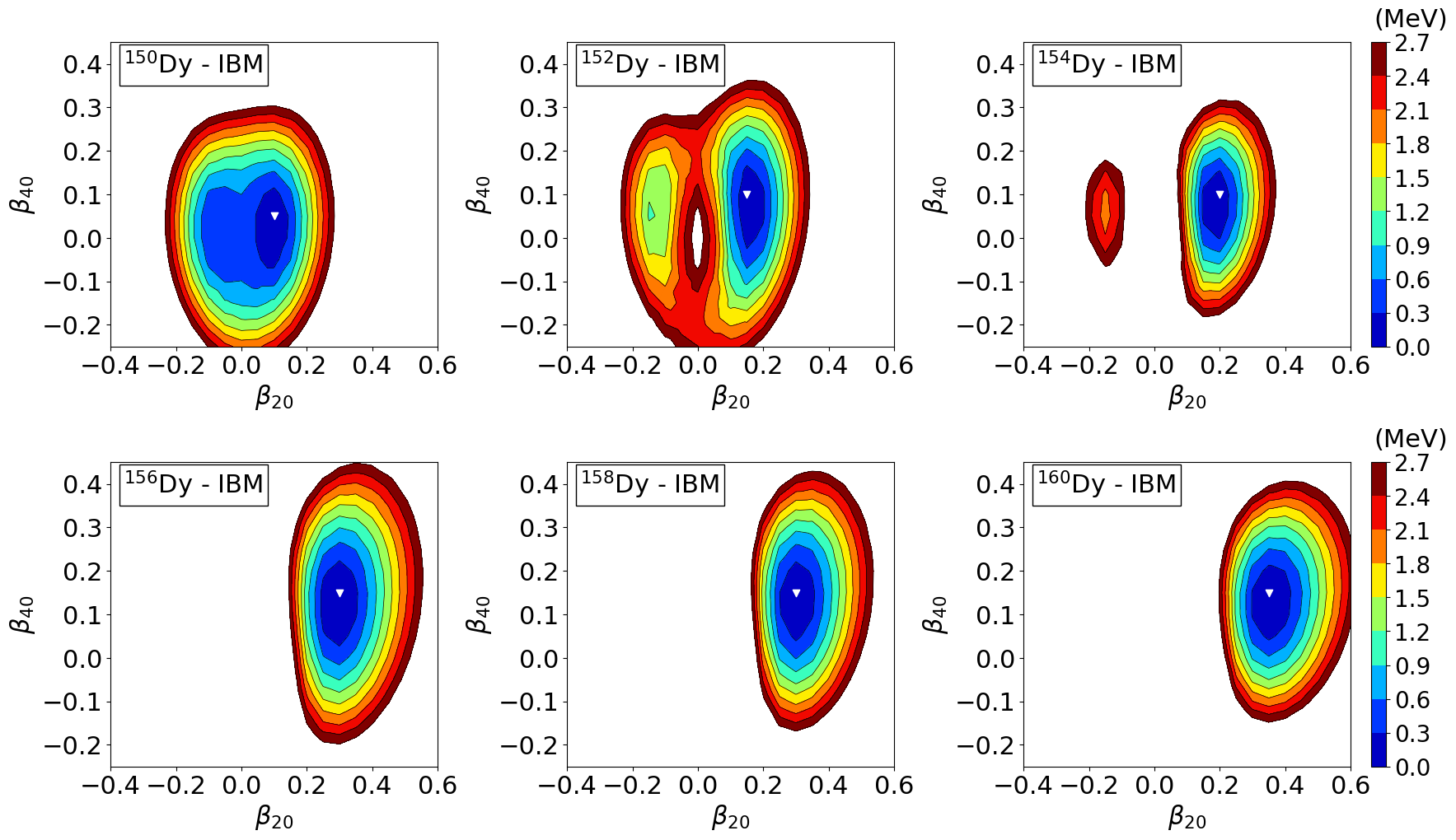}
\caption{Same as the caption for Fig. \ref{NDSCMF}, but for the mapped $sdg$-IBM energy surfaces of $^{150-160}$Dy.} 
\label{DYIBM}
\end{center}
\end{figure*}

\begin{figure*}    
\begin{center}
\includegraphics[width = 0.85\textwidth]{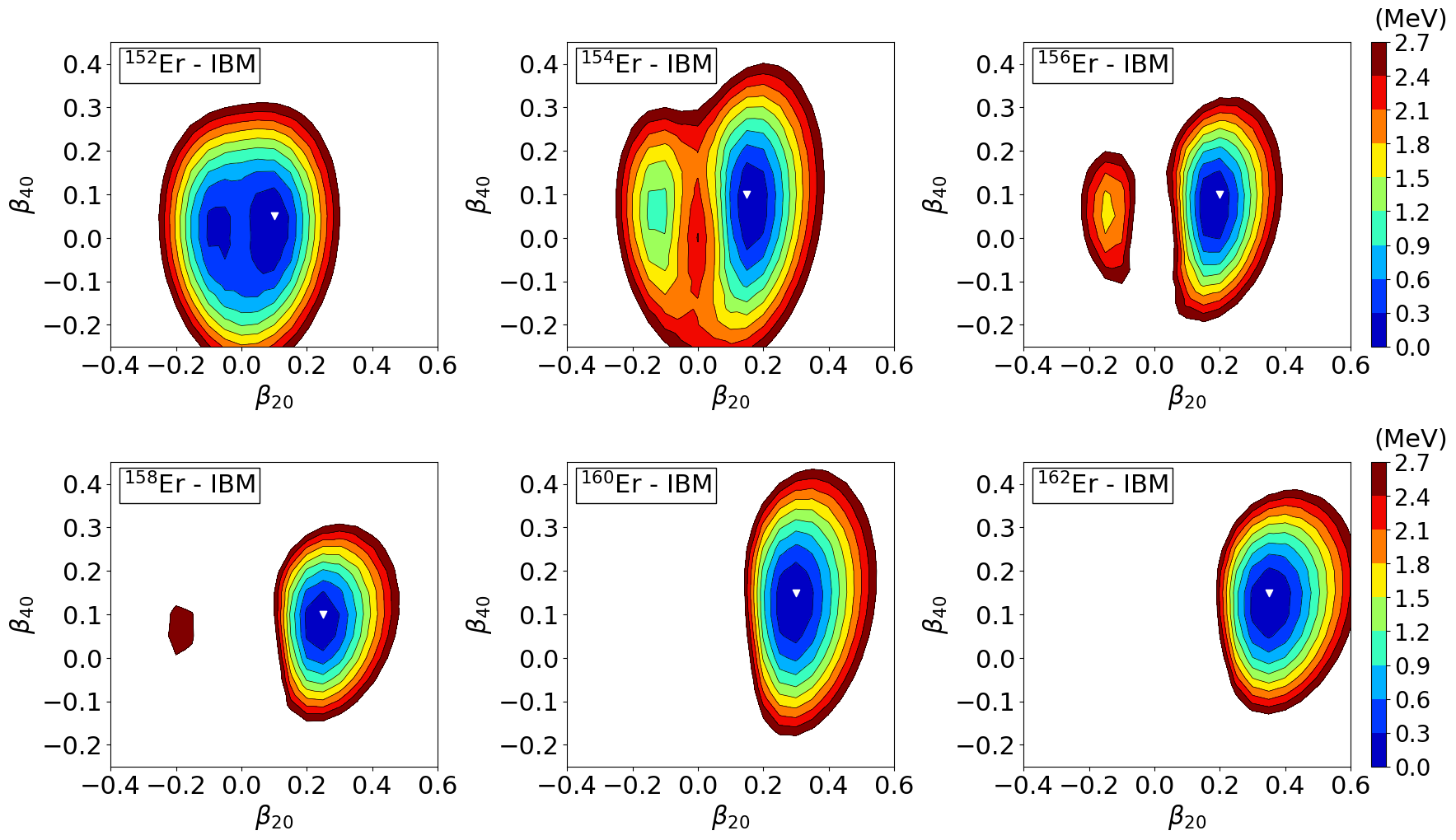}
\caption{Same as the caption for Fig. \ref{NDSCMF}, but for the mapped $sdg$-IBM energy surfaces of $^{152-162}$Er.} 
\label{ERIBM}
\end{center}
\end{figure*}

\section{Mapping the SCMF results onto the IBM space\label{sec:pes}}

Figures \ref{NDSCMF}-\ref{ERSCMF} show the PESs 
of the even-even Nd, Sm, Gd, Dy and Er isotopes 
with the neutron number within the range 
$N=84-94$, up to 2.7 MeV in energy. 
The PESs for the $N=96$ nuclei are not shown 
due to their similarity to those 
of the $N=94$ ones. 
In addition, the PESs for the Gd isotopes, 
have already been presented in Ref.~\cite{lotina2024}, 
but are depicted in Fig.~\ref{GDSCMF} for 
completeness. 
From the figures, one can notice that both the quadrupole and hexadecapole deformation parameters increase with the neutron number. The saddle point in the oblate ($\beta_{2}<0$) area is lower in energy for the $N \leq 90$ nuclei and can be seen in the PES. For heavier isotopes, the saddle point becomes higher in energy and cannot be seen in the figures. Quadrupole deformations have a similar structural evolution in all isotopes, starting from $\beta_2^{\textnormal{min}}=0.1$ at $N=84$, except for the oblate deformed $^{148}\textnormal{Gd}$ ($\beta_2^{\textnormal{min}}=-0.05$), 
with the maximum $\beta_2^{\textnormal{min}}=0.35$ calculated 
for those nuclei with $N=94$ and 96. 
It should be noted that, while $^{148}\textnormal{Gd}$ is predicted to be oblate deformed in the ground state, the PES of this nucleus shows a significant softness with respect to both quadrupole and hexadecapole deformation. The structural evolution of hexadecapole deformations is also similar in all isotopes. 
Larger hexadecapole deformations in the minimum 
are obtained for lighter nuclei, Nd and Sm, 
the largest being $\beta_4^{\textnormal{min}}=0.25$ ($^{152, 154}\textnormal{Nd}, ^{154}\textnormal{Sm}$). 
In Gd, Dy and Er isotopes, the largest 
hexadecapole deformation in the minimum is $\beta_4^{\textnormal{min}}=0.15$, present in the $N \geq 90$ region. In Dy and Er isotopes, it can be seen that the energy minima 
for the nuclei in the deformed region (with $N \geq 90$) 
become softer in the $\beta_4$ direction compared 
to those for the $N=86$ and 88 nuclei.
Earlier mean-field-type studies--e.g., 
those based on the 
axially deformed Woods-Saxon potential with 
the hexadecapole degree of freedom \cite{nazarewicz1981}, 
the total Routhian surface calculation \cite{ganioglu2014}, 
and a more recent generator coordinate method 
employing the Gogny EDF to deal with the 
axial quadrupole-hexadecapole 
coupling \cite{kumar-robledo2023}--have also found non-zero $\beta_{40}$ 
deformations in 
some rare-earth nuclei near $N=90$.

The corresponding $sdg$-IBM PESs are shown 
in Figs.~\ref{NDIBM}-\ref{ERIBM}. 
One can see that the mapping procedure reproduces some of the basic properties of the SCMF PES, such as the position of the absolute minimum and the saddle point in the $N = 84 - 90$ nuclei. 
The IBM surface is significantly larger than 
the SCMF surface, which is a general feature of the IBM due to the restricted boson space of the model. This was already discussed in the case of quadrupole - octupole mapping \cite{nomura2014}. 
The ``tail - like'' structure 
that can be seen in the SCMF PES at $N=88$ in each isotopic chain 
is also not reproduced by the IBM due to the complexities of the SCMF model, which cannot be reproduced by a simple Hamiltonian. While three-body terms would provide an improvement to the IBM PES, such terms are rarely included in the Hamiltonian and are beyond the scope of this study. In the case of the $sd$-IBM mapping, the goal was to approximately reproduce the energy as a function of the $\beta_2$ parameter, with the focus on reproducing the position of the energy minimum, the energy at $\beta_2=0$, and the saddle point in the oblate region.

\begin{figure*}    
\begin{center}
\includegraphics[width = \linewidth]{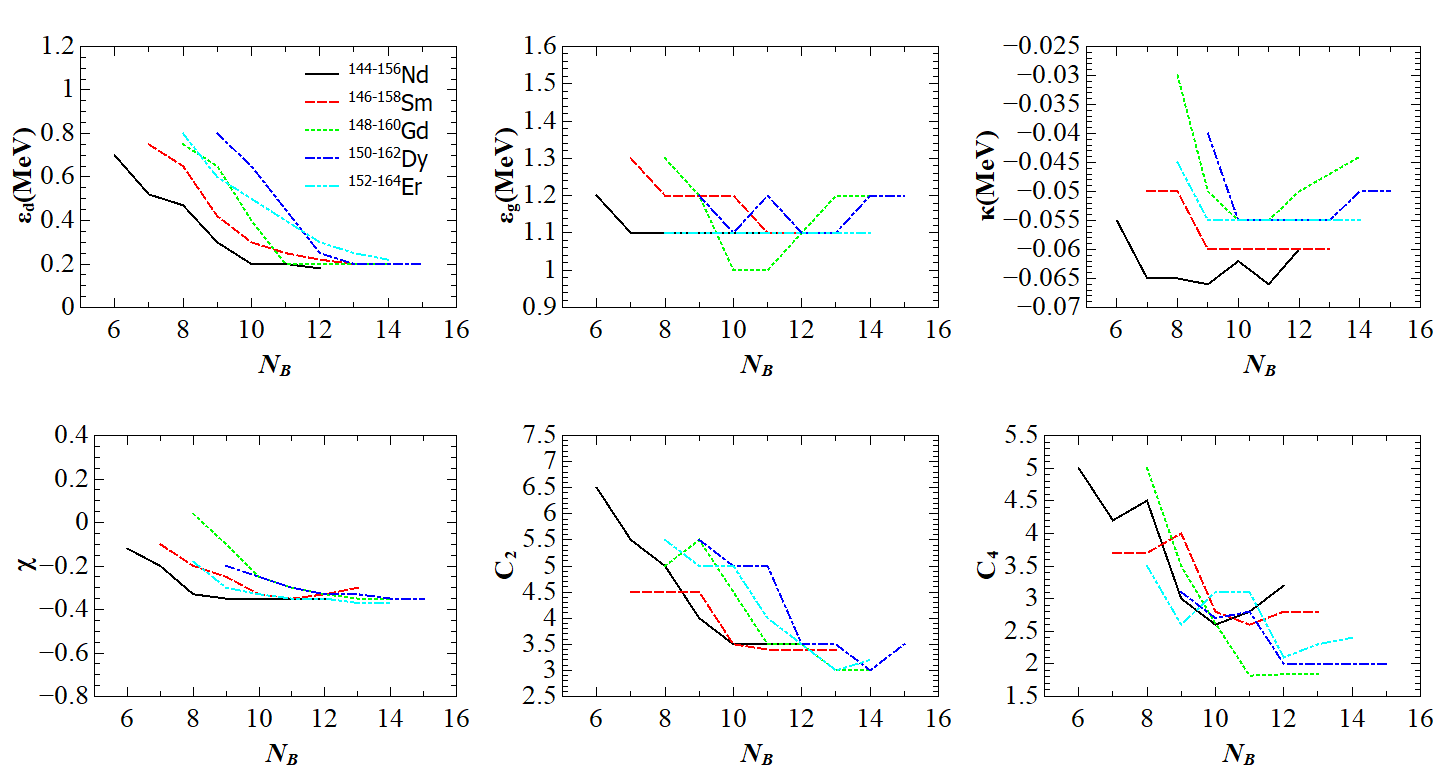}
\caption{Parameters of the $sdg$ - IBM Hamiltonian (\ref{eq5}) as functions of the boson number $N_B$.} 
\label{sdgpar}
\end{center}
\end{figure*}

\begin{figure*}    
\begin{center}
\includegraphics[width = 0.66\linewidth]{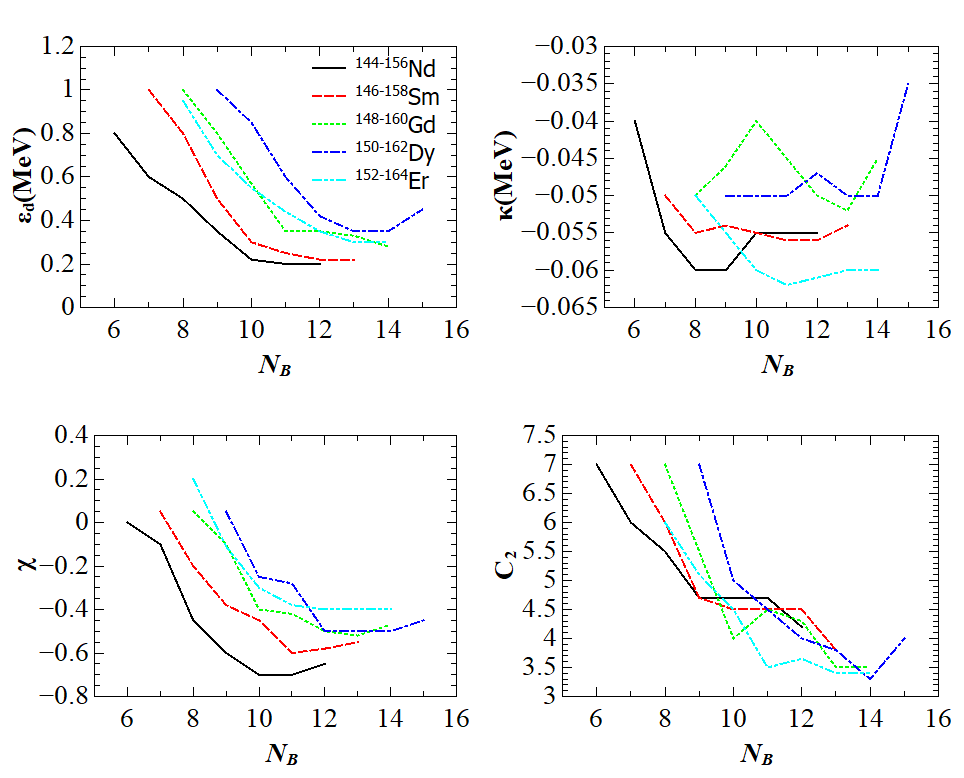}
\caption{Parameters of the $sd$ - IBM Hamiltonian (\ref{eq10}) as functions of the boson number $N_B$.} 
\label{sdpar}
\end{center}
\end{figure*}

The parameters of the $sdg$- and $sd$-IBM are shown in Figs.\ref{sdgpar} and \ref{sdpar}. The value of parameter $\sigma$ from Eq. (\ref{eq9}), not shown in Fig. \ref{sdgpar}, is set to $\sigma=3.5$ for $^{144,146}$Nd and $^{146,148}$Sm, and $\sigma=2.8$ for other Nd and Sm isotopes, 
while for Gd, Dy, and Er isotopes it is set to $\sigma=1.0$ 
[see the quadrupole operator in Eq.~(\ref{eq6})]. 
In both the $sdg$- and $sd$-IBM, the parameter $\epsilon_d$ has a maximum value in the near shell-closure region, and its value decreases as we move towards the deformed region. 
The same happens with the parameter $\chi$. 
In the $sd$-IBM, on the other hand, 
the parameter starts from a positive value in the near shell-closure region and decreases more sharply as we move into the deformed region, achieving significantly lower values from the $\chi$ parameter in the $sdg$-IBM. The $C_2$ parameter also shows similar evolution in both models. The $\kappa$ parameter in the $sdg$-IBM tends to decrease when moving to the deformed region and increase at the end of the deformed region. This is also the case in the $sd$-IBM, except in the case of Gd and Dy isotopes, for which $\kappa$ increases when moving into the deformed region. As for the parameters only present in the $sdg$-IBM, $g$ boson energy $\epsilon_g$ values fluctuate between $\epsilon_g=1.0$ and $\epsilon_g=1.3$ MeV, while the $C_4$ parameter behaves similarly to the $C_2$ parameter, the difference being that the $C_4$ parameter values tend to be smaller than the $C_2$ values for the same boson number $N_B$. 
Previous phenomenological $sdg$-IBM calculations 
on $^{152,154}$Sm have set the $g$ boson energy to be 
$\epsilon_g=$ 1.4 and 1.5 MeV, respectively \cite{kuyucak1993}, 
We note, however, that with those values we are 
not able to reproduce the desired $\beta_4^{\textnormal{min}}$ 
obtained through the SCMF calculations.

%------------------------------------------------------------------------------
%
%     Results - excitation energies and transition strengths
%
%------------------------------------------------------------------------------
 
\section{Results of the spectroscopic calculations\label{sec:results}}

In this section, we show the excitation energies 
and transition strengths. 
The computer program ARBMODEL \cite{arbmodel} is 
employed to obtain these quantities. 
The results of the $sdg$-IBM are compared with the results of the $sd$-IBM to show the effects of $g$ bosons. 
The results obtained from both models are also 
compared with the experimental data 
available in the NNDC database \cite{data}.

\subsection{Excitation energies}

\begin{figure*}    
\includegraphics[width = 0.6\linewidth]{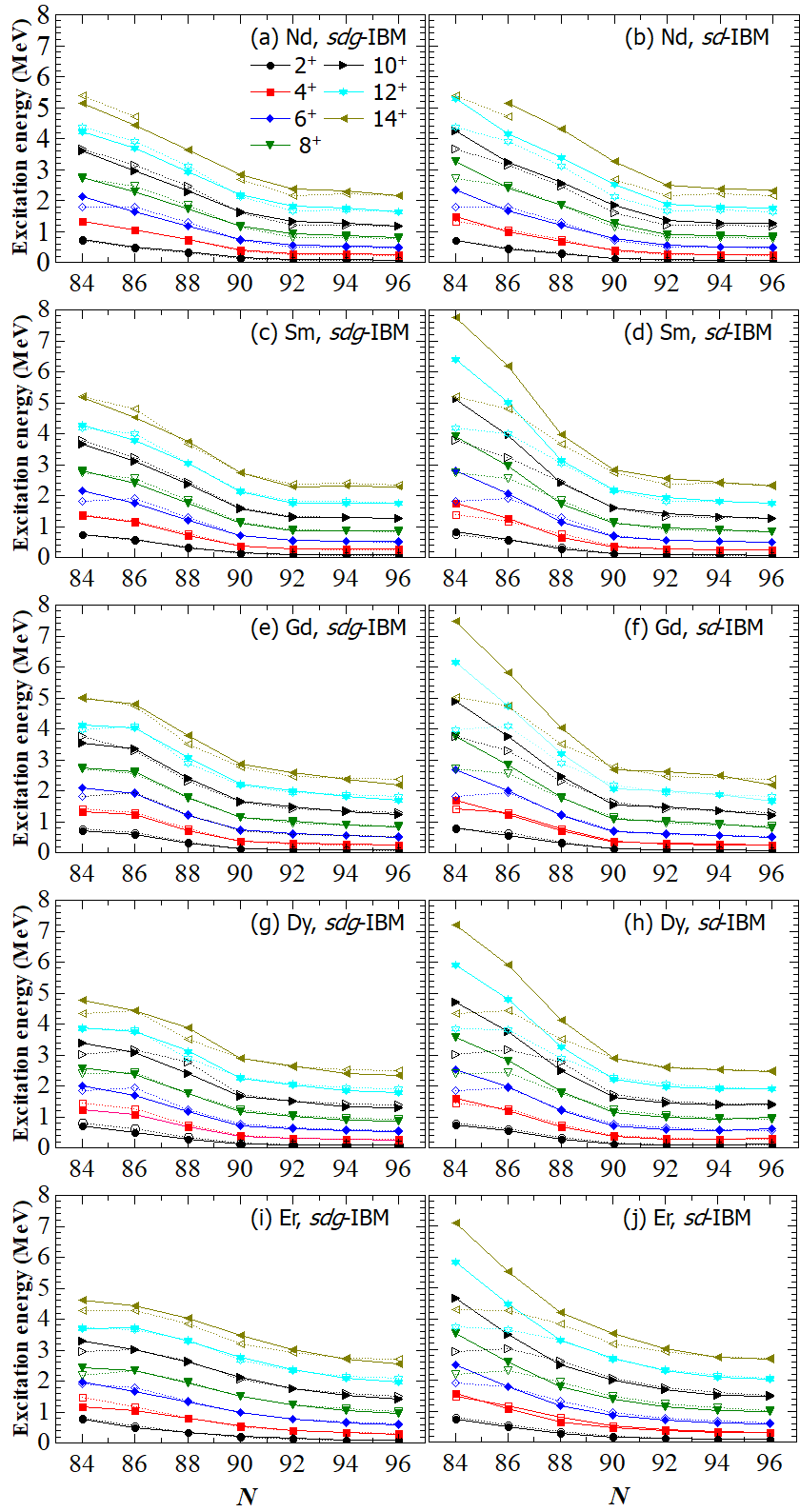}
\caption{Calculated excitation energies of the yrast band 
states up to spin $J^{\pi}=14^+$ as functions 
of the neutron number $N$ within 
the mapped $sdg$-IBM (left column) and $sd$-IBM (right column), 
represented by solid symbols connected by solid lines. 
Experimental data are taken from Ref.~\cite{data}, 
and are depicted as open symbols connected by 
dotted lines.} 
\label{yrast}
\end{figure*}

Figure \ref{yrast} shows the calculated excitation 
energies of the yrast band states with spin 
$J^{\pi}=2^+ - 14^+$. As can be seen from the figure, the $sdg$-IBM significantly improves the description of the $J^{\pi} \geq 6^+$ states in the $N \leq 88$ nuclei. 
This can be explained by looking at the 
expectation value of the $g$ boson number operator, 
which is for those states calculated to be 
$\braket{\hat{n}_g} \geq 1$. 
The energies of the yrast band states 
in the near shell-closure region are lowered 
due to the presence of $g$ bosons. 

\begin{table}[!htb]
\caption{\label{tab:R42}
Energy ratios $R_{4/2}=E_x(4^+_1)/E_x(2^+_1)$ 
for the nearly spherical nuclei with $N$=84 and 86, 
calculated with the mapped $sd$- and $sdg$-IBM, 
as compared to the experimental values \cite{data}.}
\centering
\begin{ruledtabular}
\begin{tabular}{cccc}
Nucleus & $sd$-IBM & $sdg$-IBM & Experiment \\
\hline
$^{144}$Nd & 2.11 & 1.78 & 1.89 \\
$^{146}$Nd & 2.25 & 2.05 & 2.02 \\
$^{146}$Sm & 2.12 & 1.83 & 1.85 \\
$^{148}$Sm & 2.20 & 1.98 & 2.14 \\
$^{148}$Gd & 2.13 & 1.86 & 1.81 \\
$^{150}$Gd & 2.18 & 2.15 & 2.02 \\
$^{150}$Dy & 2.15 & 1.71 & 1.81 \\
$^{152}$Dy & 2.21 & 2.16 & 2.05 \\
$^{152}$Er & 2.14 & 1.54 & 1.83 \\
$^{152}$Er & 2.24 & 2.17 & 2.07 \\
\end{tabular}
\end{ruledtabular}
\end{table}

We also summarize the energy ratios $R_{4/2}$ 
for the $N=84$ and 86 nuclei in Table \ref{tab:R42}. 
In the $N=84$ nuclei, the $sdg$-IBM predicts the ratios 
to be $R_{4/2}< 2$, which is in agreement with the experiment. This is also an effect of the $g$ boson presence, since this cannot be obtained with $sd$-IBM calculations. 
A significantly low ratio, $R_{4/2}=1.54$, 
is obtained for $^{152}$Er, 
compared to the experimental value of $R_{4/2}=1.83$. 
This is due to the fact that the $sdg$-IBM predicts 
the $4^+_1$ state 
somewhat lower in energy than the experimental value. 
This could be improved by considering the values of the parameter $\sigma$ 
to be $\sigma > 1.0$ for this nucleus. 
The calculated $R_{4/2}$ ratio that is lower 
than 2; nevertheless it agrees with experiment qualitatively, 
which, however, cannot be realized in 
the $sd$-IBM, giving $R_{4/2}=2.14>2$. 
In the $N=86$ nuclei, there is no significant 
difference between ratios obtained with the $sdg$- and $sd$-IBM. 
The two exceptions are $^{146}$Nd, where the $sdg$-IBM predicts a lower ratio, which is closer to the experimental value, and $^{148}$Sm, where the $sdg$-IBM predicts a $R_{4/2}<2$ value, which does not agree with the experiment.  

\begin{figure*}
\includegraphics[width=0.6\linewidth]{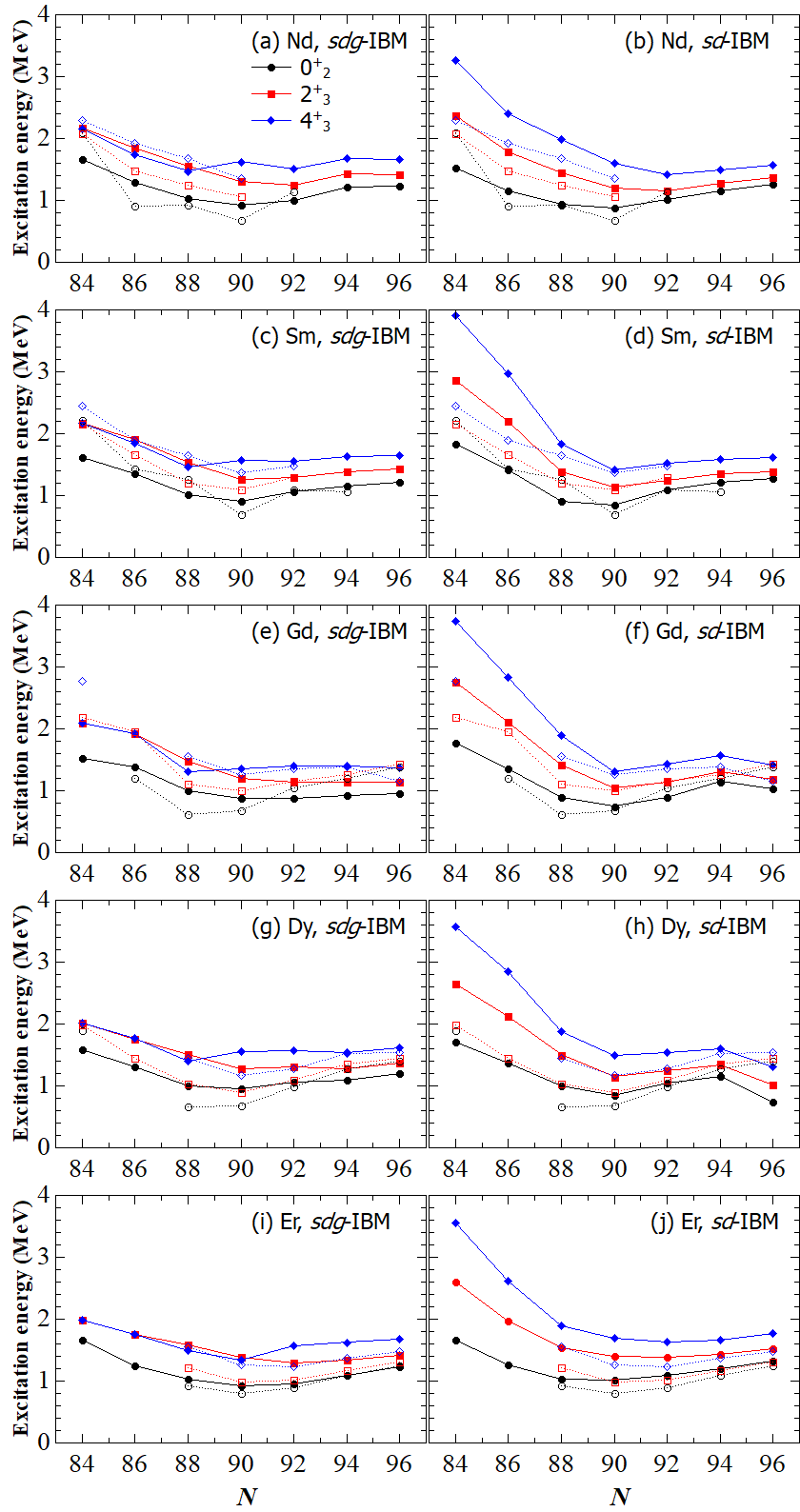}
\caption{Same as Fig.~\ref{yrast}, but for 
the $0_{2}^+$, $2_{3}^+$, and $4_{3}^+$ states.} 
\label{0+}
\end{figure*}

Figure~\ref{0+} compares the calculated 
and experimental excitation energies of 
the $0_{2}^+$, $2_{3}^+$ and $4_{3}^+$ states, 
which may be associated with the $K^{\pi}=0^+$ band 
usually present in the deformed region. 
Note that for $^{160}$Dy, the observed $0^+$ level 
at 1280 keV, 
which is suggested to be the bandhead of the first 
excited $K=0^+$ band, is shown in the plot 
[Figs.~\ref{0+}(g) and \ref{0+}(h)], 
while there are two additional excited $0^+$ 
levels at 681 and 703 keV, but 
with spin and parity not firmly established.  
As one sees in Fig.~\ref{0+}, 
the $sdg$-IBM does not provide an improved 
description of the $0_2^+$ states compared to the $sd$-IBM, 
since the expectation value of the $g$ 
boson number operator for the $0_2^+$ state 
is calculated to be $\braket{\hat{n}_g} \approx 0$. 
On the other hand, the $sdg$-IBM predicts a significantly lower $2_3^+$ and $4_3^+$ states for $N \leq 88$, which is in agreement with the experiment. 
However, in the nuclei with $N=84$ and 86, 
the two states are almost equal in energy, 
and in the $N=88$ nuclei, the $4_3^+$ state 
becomes lower in energy from the $2_3^+$ state, 
which contradicts the experiment. 
In the $N \geq 90$ deformed region, 
both $sdg$- and $sd$-IBM yield similar results. 
Overall, the $2_3^+$ and $4_3^+$ states, 
calculated by the $sdg$-IBM, are closer in 
energies to the corresponding experimental values 
in the near shell-closure region. 
The $sdg$-IBM, however, predicts the $4_3^+$ energy level 
to be so low as to be close to or even below the 
$2^+_3$ one, which does not agree with the experiment. 
The description of $0_2^+$ states is not improved in the $sdg$-IBM.

\begin{figure*}    
\includegraphics[width=0.6\linewidth]{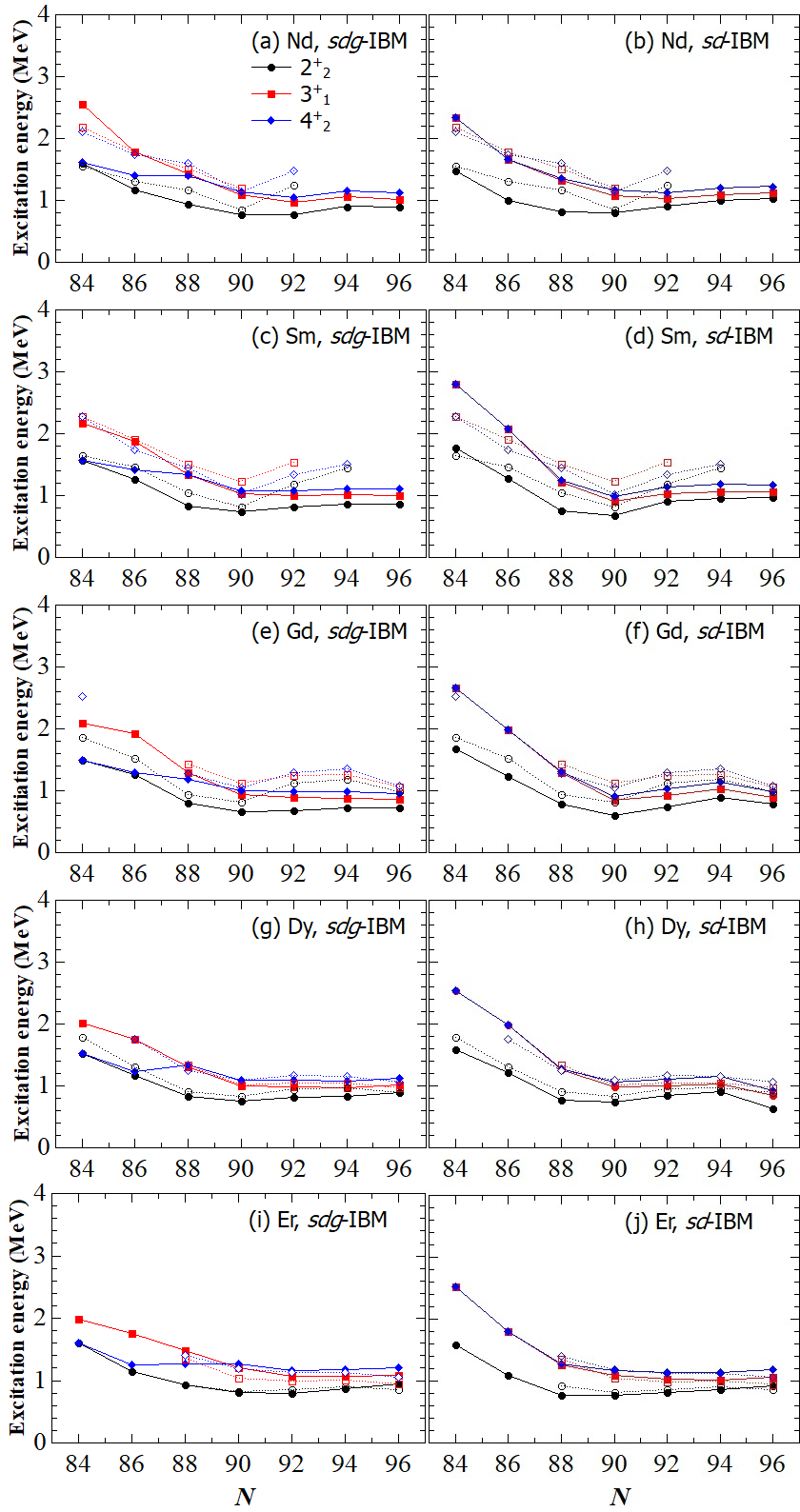}
\caption{Same as Fig.~\ref{yrast}, but for the 
$2_{2}^+$, $3_{1}^+$, and $4_{2}^+$ states.} 
\label{gamma}
\end{figure*}

Figure \ref{gamma} shows the excitation energies of the $2^+_2$, $3^+_1$, and $4^+_2$ states, associated with the $\gamma$ vibrational band. The effect of including $g$ bosons 
on the states $2_2^+$ and $3_1^+$ is minor, 
with only some small improvements in the $N \leq 88$ Nd and Sm. 
The $4_2^+$ energy level is, however, significantly 
low compared to the one obtained with the $sd$-IBM 
and to the observed level. 
In the $N=84$ nuclei, 
the $4_2^+$ state is predicted to be almost equal in energy to the $2_2^+$ state, contrary to the experiment. In the $N \geq 90$ deformed region, there are no significant differences between the $sdg$- and $sd$-IBM. The fact that the $sdg$-IBM predicts a significantly lower $4_2^+$ state compared to the experiment, points to the fact that the chosen Hamiltonian may not be suitable for the description of such states in the region near shell closures. 
A Hamiltonian with more independent parameters could potentially 
solve this problem. However, the inclusion 
of more independent parameters 
would make the mapping procedure more involved. 
We also note that the choice of the EDF, as well as 
the choice of the pairing interaction, 
affects the calculated spectrum. 
We leave those two problems for a separate study.

%------------------------------------------------------------------------------
%
% E2, E4 and E0 transitions
%
%------------------------------------------------------------------------------

\subsection{Transition strengths}
\subsubsection{Quadrupole transitions}

\begin{figure*}    
\includegraphics[width=0.6\linewidth]{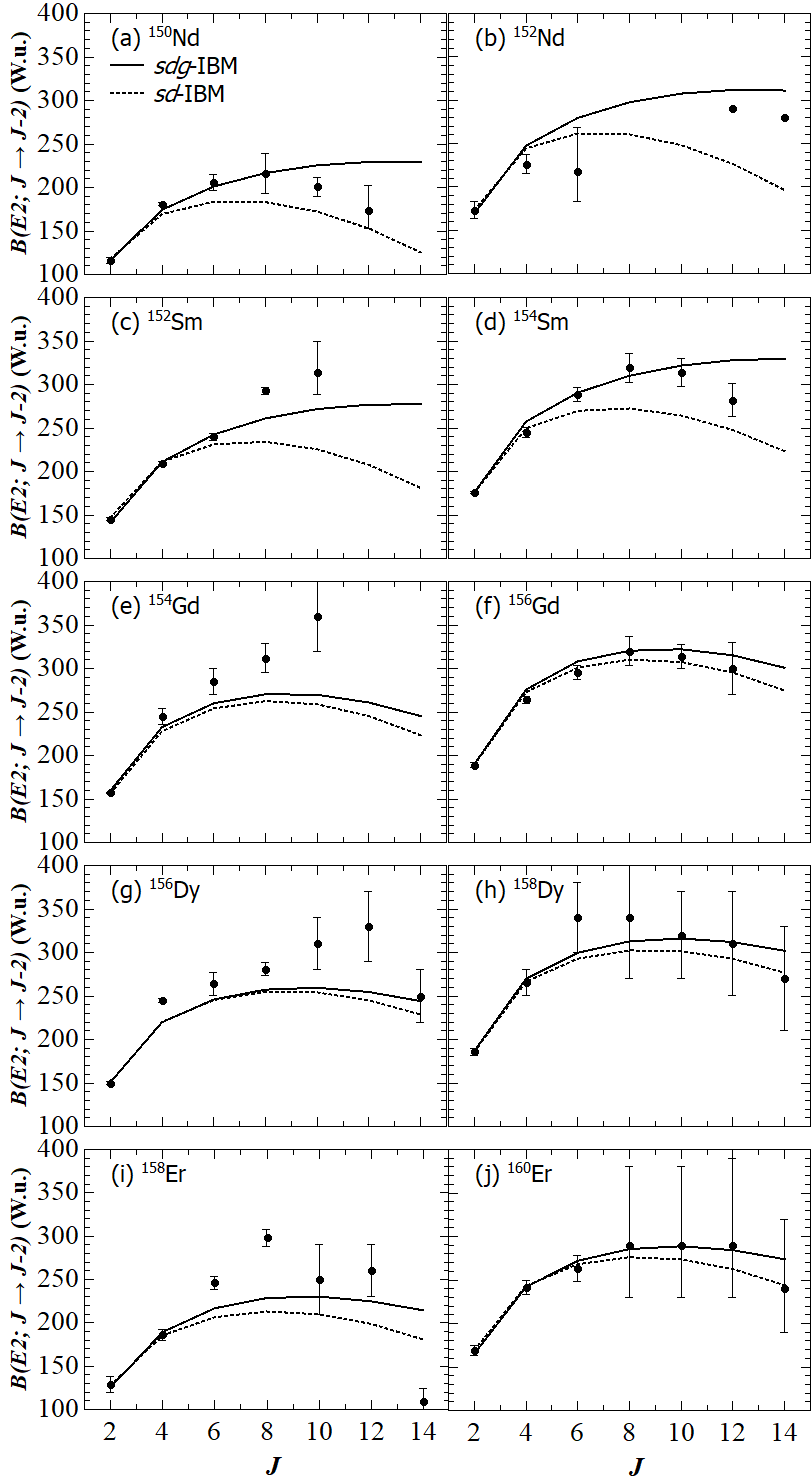}
\caption{$B(E2)$ transition strengths 
in the ground state band of the well-deformed 
$N=90$ (left) and $N=92$ (right) nuclei 
as functions of spin $J$, 
calculated with the mapped $sdg$-IBM 
(solid curves) and $sd$-IBM (dotted curves). 
The experimental data, represented by solid circles, 
are adopted from Ref.~\cite{data}.} 
\label{E2}
\end{figure*}

Figure \ref{E2} shows the $B(E2; J \rightarrow J-2)$ transition strengths in the ground state bands of the well deformed $N=90$ and 92 isotopes. We consider these isotopes due to the fact that most of the data on $E2$ transitions 
are available for these isotopes, 
which makes them the ideal cases to examine when comparing the $E2$ transition strengths between the $sdg$- and $sd$-IBM. The effective charges $e_2^{sdg,sd}$ are fitted to reproduce the experimental data on the first $B(E2; 2^+ \rightarrow 0^+)$ transition \cite{data}. One can notice a significant difference between the behavior of the ground state band $E2$ transitions in Nd and Sm isotopes, compared to heavier ones. In Nd and Sm, the predicted transition strengths for states $J^{\pi} \geq 6^+$ are significantly larger than the $sd$-IBM calculated transitions, which is not the case in other isotopes. This can be explained by the fact that in those isotopes 
the $(d^{\dagger} \times \tilde{g} + g^{\dagger} \times \tilde{d})^{(2)}$ 
term of the quadrupole operator $\hat Q^{(2)}$ 
contributes more to the calculated transitions due 
to the larger values of the parameter $\sigma > 1.0$. 
The calculated transition strengths, especially in $^{152,154}$Sm, seem to correspond to the axial rotor calculations \cite{kuyucak1993, kuyucak1994}. It can be seen that the $sdg$-IBM in the shown Nd and Sm isotopes improves the results of the $B(E2; J \rightarrow J-2)$ strengths for $J=6^+, 8^+, 10^+$. In Gd, Dy, and Er isotopes, the $sdg$-IBM only slightly increases the $E2$ transition strengths from $J^{\pi} \geq 6^+$ states compared to the $sd$-IBM, which can be attributed to the fact that the value of the parameter $\sigma=1.0$ 
is chosen. 
At $N=90$, both models underestimate 
the measured transition strengths, while at $N=92$, both models reproduce the measured strengths well. Due to the fact that the margins of error are quite large in $N=92$ Dy and Er isotopes, it cannot be concluded whether the $sdg$-IBM improves the description of higher $E2$ transition strengths in those isotopes.

%-----------------------------------------------------------
%
%       E4 transitions
%
%-----------------------------------------------------------

\subsubsection{Hexadecapole transitions}

\begin{figure*}    
\includegraphics[width=0.7\linewidth]{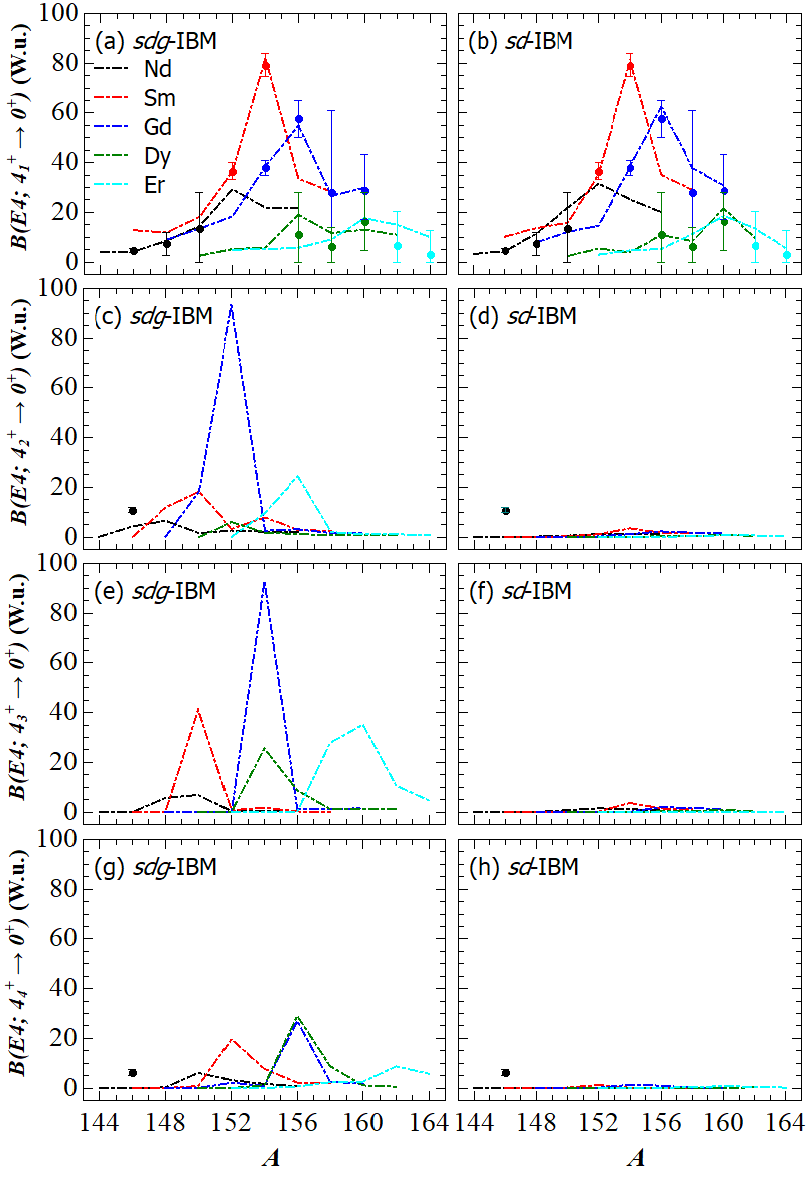}
\caption{
$B(E4)$ strengths in W.u. for the 
transitions of 
the first [panels (a) and (b)], 
second [panels (c) and (d)], 
third [panels (e) and (f)], and 
fourth [panels (g) and (h)] 
$4^+$ state to the $0^+_1$ ground state 
as functions 
of the mass number $A$, calculated with 
the mapped $sdg$-IBM (left column) 
and $sd$-IBM (right column). 
Experimental data are taken from 
Refs.~\cite{data, ronningen1977, wollersheim1977, ronningen21977}, 
and are indicated by solid circles in the plots.} 
\label{E4}
\end{figure*}

Figure~\ref{E4} shows the 
$B(E4; 4_n^+ \rightarrow 0_1^+)$ ($n=1,2,3,4$)
transition strengths. 
The $e_4^{sdg, sd}$ effective charges are 
fitted to experimental data on the 
$B(E4; 4_1^+ \rightarrow 0_1^+)$ from the first $4^+$ state to the ground state \cite{data, ronningen1977, wollersheim1977, ronningen21977}. For isotopes with no available experimental data, effective charge values are chosen so that they start from lower values, peak near $N=92$, 
and then decrease again. 
These transition strengths are shown in Figs.~\ref{E4}(a) and \ref{E4}(b). 
In Figs.~\ref{E4}(c)-\ref{E4}(h), $E4$ transition strengths from higher $4^+$ states are shown. The $sdg$-IBM predicts several large $E4$ transition strengths from these states in certain isotopes, which is expected in the case of hexadecapole deformed nuclei with a $K=4^+$ band. The $sd$-IBM predicts all of these transition strengths to vanish, which points to the necessity of considering the $g$ boson in the description of $E4$ transitions from higher $4^+$ states. Unfortunately, due to the lack of experimental data on these $E4$ transitions, it is not possible to see how well the mapped $sdg$-IBM predicts the values of these transition strengths.

%----------------------------------------------------------------------------
%
%    E0 transitions
%
%----------------------------------------------------------------------------

\subsubsection{Monopole transitions}

\begin{figure*}    
\includegraphics[width=0.8\linewidth]{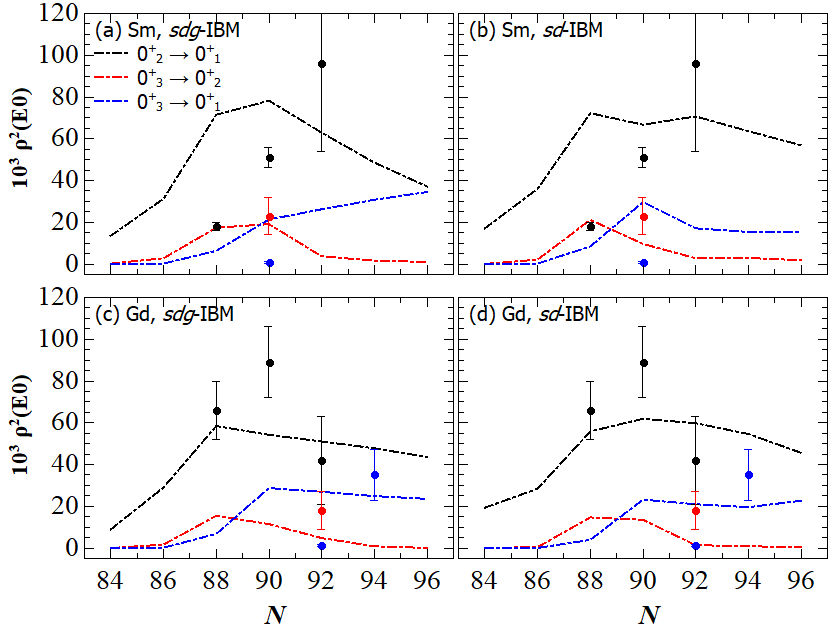}
\caption{$\rho^{2}(E0; 0_i^+ \rightarrow 0_j^+)$ 
values as functions of the neutron number $N$ 
for Sm and Gd isotopes, calculated with 
the mapped $sdg$-IBM (left column) 
and $sd$-IBM (right column). 
Experimental values are adopted from 
Refs.~\cite{data,kibedi2005}, and 
are plotted as solid circles.} 
\label{E0}
\end{figure*}

Figure \ref{E0} shows the monopole strengths $\rho^{2}(E0; 0_i^+ \rightarrow 0_j^+)$, with $i=2,3$ and $j=1,2$, for isotopes of Sm and Gd. We choose to show these isotopes, since the experimental data on monopole strengths is only available for these isotopes \cite{data,kibedi2005}. The choice of $\eta = \gamma = 0.75 \: \textnormal{fm}^2$ is made to reproduce most of the available experimental data. The $sdg$-IBM does not significantly improve the calculated monopole strengths compared to the $sd$-IBM. Both models overestimate the strengths of $0_2^+ \rightarrow 0_1^+$ transitions in $^{150,152}$Sm and underestimate the same strength in $^{154}$Gd. The calculated strengths of $^{154}$Sm and $^{152,156}$Gd are within the margins of error of the measured strengths. The $sdg$-IBM does slightly improve the description of the $0_3^+ \rightarrow 0_2^+$ transition in $^{154}$Sm and the $0_3^+ \rightarrow 0_1^+$ transition in the $^{158}$Gd. Overall, the $sdg$-IBM does not differ significantly from the $sd$-IBM in the description of the monopole strengths, which is expected, since the $sdg$-IBM calculations do not predict a presence of $g$ bosons in $0^+$ states up to $0_3^+$. 
For example, in $^{154}$Sm, the lowest $0^+$ 
state that contains one $g$ boson, with 
the expectation value $\braket{\hat{n}_g} \approx 1$, 
is the $0_5^+$ state. 
In principle, it is possible to fit $\eta$ 
and $\gamma$ separately for each isotope. 
However, since our goal was to see the effect of 
$g$ bosons in monopole transitions, 
we follow the method of \cite{vanisacker2012} 
and set fixed values of $\eta$ and $\gamma$ parameters.

\section{Summary\label{sec:summary}}

We have shown an extended analysis 
of the impact of hexadecapole deformations 
on the excitation energy spectra and transition 
strengths in even-even rare-earth nuclei, 
ranging from the near spherical to the well deformed ones. 
The quadrupole-hexadecapole constrained SCMF 
PES has been mapped onto the corresponding 
PES of the IBM, and this procedure completely 
determines the parameters of the $sdg$-IBM 
Hamiltonian, based on the microscopic calculations. 
The inclusion of $g$ bosons has a significant effect on $J^{\pi} \geq 6^+$ yrast states in the $N \leq 88$ nuclei near neutron magic number $N=82$. The mapped $sdg$-IBM lowers the energies of aforementioned 
states to agree with the observed spectra. 
In the case of non-yrast states and corresponding bands, the $sd$-IBM seems to be sufficient in the description of such states, with the $sdg$-IBM making only a minor contribution, e.g., $2_3^+$ and $4_3^+$ states of the $K^{\pi}=0^+$ band in the $N=84$ and 86 nuclei.
As for the transitions, in the well deformed nuclei with $N=90$ and 92, the $sdg$-IBM calculation yields higher $B(E2; J \rightarrow J-2)$ values for $J^{\pi} \geq 6^+$ yrast states, which does seem to be an improvement of the results, especially in the case of $^{150,152}$Nd and $^{152,154}$Sm. In the case of monopole transitions between $0^+$ states, the effect of the $g$ boson seems to be minor. 
In the well deformed region, the $sdg$-IBM predicts the existence of the $K^{\pi}=4^+$ band with an enhanced $B(E4; 4^+ \rightarrow 0^+)$ hexadecapole transition to the ground states. The fact that the $sd$-IBM cannot predict larger hexadecapole transition strengths from higher $4^+$ states points to a necessity of including the $g$ boson in the description of the hexadecapole transitions. Unfortunately, due to the lack of experimental data on such transitions, it is not possible to see how well the $sdg$-IBM reproduces such transitions.
Now that we have shown the usefulness of the mapped $sdg$-IBM, we can expand our study to the even-odd and odd-odd rare-earth nuclei, as well as extend our model to the more complex $sdg$-IBM-2 to study properties such as scissors modes in rare-earth nuclei. It could also be interesting to systematically study how sensitive the parameters are to the choice of the EDF in the SCMF calculations.

\acknowledgments
The work of L.L. is financed within 
the Tenure Track Pilot Programme of 
the Croatian Science Foundation and 
the \'Ecole Polytechnique F\'ed\'erale de Lausanne, 
and the project TTP-2018-07-3554 
Exotic Nuclear Structure and Dynamics, 
with funds from the Croatian-Swiss Research Programme. 

\bibliography{refs}

\end{document}